\documentclass[runningheads]{llncs}
\usepackage{graphicx}

\usepackage{tikz}
\usepackage{comment}
\usepackage{amsmath,amssymb} %
\usepackage{color}

\usepackage[accsupp]{axessibility}  %
\usepackage{booktabs}
\usepackage{multirow}

\usepackage{amsmath}

\newcommand*{\tran}{^{\mkern-1.5mu\mathsf{T}}}

\DeclareMathAlphabet{\mbf}{OT1}{ptm}{b}{n}
\newcommand{\mbs}[1]{{\boldsymbol{#1}}}

\usepackage{xspace}
\makeatletter
\DeclareRobustCommand\onedot{\futurelet\@let@token\@onedot}
\def\@onedot{\ifx\@let@token.\else.\null\fi\xspace}

\def\eg{\emph{e.g}\onedot} 
\def\ie{\emph{i.e}\onedot}

\def\etal{\emph{et al}\onedot}
\makeatother

\usepackage[pagebackref=true,breaklinks=true,colorlinks,bookmarks=false]{hyperref}

\begin{document}
\pagestyle{headings}
\mainmatter
\def\ECCVSubNumber{6573}  %

\title{NOCT: Nonlinear Observability with Constraints and Time Offset} %

\titlerunning{NOCT}
\author{Jianzhu Huai\inst{1} \and%
Yukai Lin\inst{2} %
\and Yujia Zhang\inst{3}%
}
\authorrunning{J. Huai et al.}
\institute{Wuhan University, Wuhan Hubei 430079, China
\email{jianzhu.huai@whu.edu.cn} \and
ETH, Zurich Switzerland
\email{linyukai@outlook.com} \and
York University, Toronto ON, Canada
\email{zhang@gmail.com}
}
\maketitle

\begin{abstract}
Nonlinear systems of affine control inputs overarch many sensor fusion instances. Analyzing whether a state variable in such a nonlinear system can be estimated (i.e., observability) informs better estimator design.
Among the research on local observability of nonlinear systems, approaches based on differential geometry have attracted much attention for the solid theoretic foundation and suitability to automated deduction.
Such approaches usually work with a system model of unconstrained control inputs and assume that the control inputs and observation outputs are timestamped by the same clock.
To our knowledge, it has not been shown how to conduct the observability analysis with additional constraints enforced on the system's observations or control inputs.
To this end, we propose procedures to convert a system model of affine control inputs with linear constraints into a constraint-free standard model which is apt to be analyzed by the classic observability analysis procedure.
Then, the whole analysis procedure is illustrated by applying to the well-studied visual inertial odometry (VIO) system which estimates the camera-IMU relative pose and time offset.
The findings about unobservable variables under degenerate motion concur with those obtained with linearized VIO systems in other studies, whereas the findings about observability of time offset extend those in previous studies.
These findings are further validated by simulation.

\end{abstract}

\section{Introduction}
Being prevalent in robots and augmented reality, state estimation usually integrates data from sensors of different modalities mounted on a platform to estimate the state of the platform, \eg, position and orientation.
Examples of state estimation include calibration, odometry, and mapping.
For state estimation, identifying unobservable state variables and handling them properly in estimation is crucial to designing a good estimator \cite{isidoriNonlinear2013}.
Given a multi-sensor system, identifying unobservable state variables without using real data is usually achieved by observability analysis.

Much research has been conducted on this topic.
For a linear system, it is fairly easy to analyze its observability by using the Kalman's test \cite{kalmanGeneral1960}.
This gets much harder for nonlinear systems for which conclusions drawn by the observability analysis usually are confined to a local area in the state-space, leading to the ``local weak observability'' \cite{hermannNonlinearControllabilityObservability1977}.
One approach is to linearize the system and assume that the control inputs are piece-wise constant \cite{goshen-meskinObservability1992}.
This approach often involves complex integrals and 
requires human experience to identify the unobservable directions, thus it is time-consuming to verify the results.
Another problem of this approach is that we do not know for certain whether all unobservable directions are identified.

An alternative approach is to directly work with the differential equations of the system and
analyze its observability with Lie derivatives as proposed in \cite{hermannNonlinearControllabilityObservability1977}.
The procedure in \cite{hermannNonlinearControllabilityObservability1977} has been extended to systems where the inputs directly affect the output observations \cite{villaverdeInputDependent2019},
and to systems where some inputs are unknown \cite{martinelliNonlinearUnknownInput2022}.
These approaches provide analytical solutions and are suitable for automatic deduction.

However, existing approaches based on Lie derivatives assume that the control inputs to the system are unconstrained (general) and that there is a zero time offset between the clock to timestamp control inputs and the clock to timestamp the observation outputs.
To our knowledge, it has not been shown how to use the method based on Lie derivatives to analyze a nonlinear system when additional constraints are added.
As an example, for a visual inertial odometry (VIO) system which estimates platform motion by fusing camera images and inertial measurement unit (IMU) data, it often face scenes where the landmarks are on a plane or times when the platform is restricted to move on a plane.

This paper presents several procedures to convert a nonlinear system of affine inputs with linear constraints to a nonlinear system of affine inputs (termed a standard nonlinear system), so that the observability properties of the original system can be obtained by analyzing the converted system with a well-founded automated procedure.
Then, for a monocular VIO problem where motion of the platform, camera-IMU extrinsic parameters and time offset are to be estimated, we analyze its observability properties under three types of constraints.
The identified unobservable variables in these cases agree with the findings reported in other studies.
These findings are also validated in simulation with a filter-based VIO method, showing the effectiveness of proposed conversion procedures.
As for the time offset, we supplement our earlier work \cite{huaiObservabilityAnalysisKeyframebased2022} which identifies the condition when the time offset is unobservable with simulation results.

In summary, the contributions of this paper are:
\begin{enumerate}
\item To our knowledge, we are the first to deal with observalibity of nonlinear systems of affine control inputs with linear constraints by using the Lie derivatives. For this problem, we propose conversion procedures suitable to automation.

\item The effectiveness of the proposed method is demonstrated on a VIO system with on-the-fly extrinsic and temporal calibration and confirmed by simulation.
\end{enumerate}

\section{Related Work}
This section briefly reviews recent research in observability analysis.
The notion of observability was proposed by Kalman \cite{kalmanGeneral1960} for analyzing linear systems.
Hermann and Krener \cite{hermannNonlinearControllabilityObservability1977} extended observability analysis to nonlinear systems by using differential geometry methods.
In contrast to linear systems, the nonlinear system's observability analysis is typically local, \ie, only valid for a neighborhood in the state-space.
There are also global observability analysis methods for a nonlinear system, \eg, \cite{tangINS2009}, but these problem-specific techniques do not easily adapt to other nonlinear systems.

Two streams of local observability analysis methods are popular for nonlinear systems, discrete model analysis for piece-wise constant systems,
\eg, \cite{goshen-meskinObservability1992},
and continuous model analysis based on the differential equations of a system, \eg, \cite{martinelliVisualinertialStructureMotion2013}. The discrete model analysis works with the observability matrix of the linearized system, whereas the continuous model analysis works with the observability matrix derived from Lie derivatives of the continuous system.

Using the observability matrix built with piece-wise constant transition functions,
Li and Mourikis \cite{liHighprecisionConsistentEKFbased2013} improved consistency of a filter-based VIO algorithm by maintaining observability properties of VIO.
Recently, \cite{yangOnlineSelfcalibrationVisualinertial2022} analyzed the unobservable directions of a linearized VIO system with full self-calibration under motion constraints.
Li and Stueckler \cite{li_visual-inertial_2022} incorporated constraints due to velocity controls and planar motion as observations into the VIO system and 
analyzed observability of state variables by using the linearized system model.
Generally speaking, the linearized model requires discretizing the differential equations and brings about 
complex integrals which are difficult to deal with in automatic symbolic deduction.
By manually identifying the unobservable directions, this approach is likely to miss some unobservable directions. For instance, consider the observability properties obtained in \cite{yangOnlineSelfcalibrationVisualinertial2022} based on a linearized VIO model, the camera-IMU system would have a different number of unobservable calibration parameters when the axis labels of the IMU are swapped (\eg, x-y-z to z-x-y), indicating that some unobservable parameters are missed out by manual checking.

Much work has been done on observability analysis with continuous models, \cite{mirzaeiKalman2008}, \cite{kellyVisualinertialSensorFusion2011}, \cite{heschCameraIMUbasedLocalizationObservability2014},
\cite{jungConstrained2020},
\cite{villaverdeFull2019},  \cite{huaiObservabilityAnalysisKeyframebased2022}, and \cite{li_visual-inertial_2022}.
Mirzaei \etal \cite{mirzaeiKalman2008} applied the continuous analysis to extrinsic calibration of a camera-IMU system based on an Extended Kalman filter (EKF).
Kelly \etal \cite{kellyVisualinertialSensorFusion2011} proved that the camera-IMU extrinsic parameters are observable with natural features.
Jung \etal \cite{jungConstrained2020} analyzed observability of an EKF-based estimator whose state vector includes the IMU intrinsic parameters.
Villaverde \cite{villaverdeFull2019} extended the continuous model analysis approach to systems where the control directly feeds to the output (\ie, direct feed-through).
The studies in \cite{maesObservability2019} and \cite{martinelliNonlinearUnknownInput2022} extended the continuous analysis to systems with unknown inputs 
assuming the control input is piece-wise constant.
Huai \cite{huaiObservabilityAnalysisKeyframebased2022} showed that full self-calibration of the camera-IMU system could be achieved under general motion by using differential equations of the VIO system.

These approaches working with continuous models have not explicitly considered systems with constraints on observations or control inputs.

Previous approaches for temporal calibration identify conditions when the time offset is unobservable by using system-specific derivation, \eg, \cite{liOnlineTemporalCalibration2014}.
Degenerate cases for a lidar-IMU system with on-the-fly extrinsic and temporal calibration were identified in \cite{zuo_lic-fusion_2020} by using a linearized model.
For a camera-IMU system, Yang \etal \cite{yangOnlineSelfcalibrationVisualinertial2022} analyzed cases where the time offset is unobservable by using a linearized system model.
Recently, in \cite{huaiObservabilityAnalysisKeyframebased2022}, degenerate cases for time offset estimation were identified for a nonlinear system of affine inputs, encompassing cases reported in \cite{zuo_lic-fusion_2020,yangOnlineSelfcalibrationVisualinertial2022}.
This paper provides additional simulation results to supplement the findings of \cite{huaiObservabilityAnalysisKeyframebased2022}.

\section{Nonlinear Observability Analysis Background}
This section briefly presents the nonlinear observablity analysis procedure with Lie derivatives developed in \cite{hermannNonlinearControllabilityObservability1977} 
and elaborated in \cite{martinelliObservability2020}, which is the basis for the observability analysis with constraints.

The two popular observability analysis approaches for nonlinear systems, one based on the discrete model,
and the other based on the continuous model, are briefly described in order for a better perspective.
The given information to the analysis includes the observations and control inputs in a time interval $\mathcal{I} = [t_0, t_0 + \Delta t]$.

The method based on the discrete model first linearizes the differential equations of the system around states at different time points in $\mathcal{I}$, obtaining
the discrete transition model and observation model,
and finally by using an observability matrix built from these models, considers the condition to solve for the first of those states for the time point sequence.
Usually, the observability matrix involves integrals of control inputs.

In contrast, the method based on differential geometry works with time derivatives of the process model and the observation model of the nonlinear system.
The values of these time derivatives are known given the set of observations in $\mathcal{I}$ according to the Taylor expansion of the observation model at the start time point.
These time derivatives are a set of functions of the state at the start time point and the control inputs.
By the inverse function theorem, if the gradient of the set of equations is full rank,
then the initial state is locally observable.
Since the set of equations is in fact the product of a matrix involving only the control inputs 
and a vector involving only Lie derivatives of the observation equations $\mbf h$.
Therefore, the gradient of the set of equations can be equivalently represented by the gradient of the Lie derivatives of $\mbf h$ (hence the nonlinear observability matrix in \eqref{eq:obs-matrix}),
assuming that the control inputs are unconstrained (\ie, can be chosen by a user so that the maximum rank of the observability matrix is achieved).

In summary, Table~\ref{tab:lin-vs-nonlin} lists differences between observability analysis based on the discrete model and that based on the continuous model.

\begin{table}[!htb]
\centering
\caption{The differences between observability analysis with the linearized system model and that with the Lie derivatives of the nonlinear system.}
\label{tab:lin-vs-nonlin}
\begin{tabular}{@{}lll@{}}
\toprule
\textbf{Differences}                                                                         & \textbf{Linearized model}                                                                                   & \textbf{\begin{tabular}[c]{@{}l@{}}Nonlinear \\ continuous model\end{tabular}}                   \\ \midrule
Assumptions                                                                                   & \begin{tabular}[c]{@{}l@{}}A series of state\\ vectors is roughly known;\\ First order approx.\end{tabular} & \begin{tabular}[c]{@{}l@{}}One state vector\\ is roughly known;\\ Inputs are general.\end{tabular} \\ \midrule
\begin{tabular}[c]{@{}l@{}}Observablity matrix \\ $\mathcal{O}$ involves inputs?\end{tabular} & \begin{tabular}[c]{@{}l@{}}Integrals of \\ inputs are in $\mathcal{O}$.\end{tabular}                        & \begin{tabular}[c]{@{}l@{}}Inputs are \\ not in $\mathcal{O}$.\end{tabular}                        \\ \midrule
\begin{tabular}[c]{@{}l@{}}Amenable to \\ automated deduction\end{tabular}                   & Hardly                                                                                     & Yes                                                                              \\ \midrule
\begin{tabular}[c]{@{}l@{}}Identify the\\ complete null space\end{tabular}                    & \begin{tabular}[c]{@{}l@{}}Very difficult to\\ manually find all.\end{tabular}                              & \begin{tabular}[c]{@{}l@{}}If the automated\\ program is efficient.\end{tabular}                   \\ \bottomrule
\end{tabular}
\end{table}

Now we briefly present fundamentals of the nonlinear observability analysis.
According to \cite{martinelliObservability2020}, a state at time $t_0$, $\mbf x(t_0)$, is weakly observable if there is a neighborhood in which all its neighbors can be
distinguished from itself by the knowledge of outputs and unconstrained inputs in a time interval $\mathcal{I} = [t_0, t_0 + \Delta t]$.
For a noise-free system that is affine in the control inputs $\mbf u_i$, $i=1, 2, \cdots, m$,
\begin{equation}
	\label{eq:affine-input-sys}
	\dot{\mbf{x}} = \mbf{f}_0(\mbf x) + \sum_{i = 1}^{m}\mbf f_i(\mbf x) u_i, \quad 
	\mbf{y} = \mbf{h(x)}
\end{equation}
the sufficient condition for weak observability of its state $\mbf x$ is that the observability matrix built from the outputs $\mbf h(\mbf x)$ has full rank
\cite{hermannNonlinearControllabilityObservability1977}, \cite{martinelliObservability2020}.
The observability matrix $\mathcal{O}$ consists of gradients of Lie derivatives of
$\mbf h(\mbf x)$ along vector fields $\mbf f_i$ of the control inputs $u_i$.
For a vector output function $\mbf h$, its first order Lie derivative along the vector field
 $\mbf f_i$ of one column is defined by
\begin{equation}
	\mathcal{L}^1_{\mbf f_i} \mbf{h} = \nabla \mbf h \cdot \mbf f_{i}
\end{equation}
The zeroth order Lie derivative is defined as
\begin{equation}
	\mathcal{L}^0 \mbf h = \mbf h.
\end{equation}
Higher order Lie derivatives can be computed recursively by 
\begin{equation}
	\mathcal{L}^2_{\mbf f_i, \mbf f_j} \mbf{h} = \mathcal{L}^1_{\mbf f_j} \mathcal{L}^1_{\mbf f_i} \mbf{h}.
\end{equation}
With these Lie derivatives of the observation function, the entire observability matrix $\mathcal{O}$ is given by
\begin{equation}
\label{eq:obs-matrix}
	\mathcal{O} = \begin{bmatrix}
		\nabla \mathcal{L}^0 \mbf h \\
		\nabla \mathcal{L}^1_{\mbf f_i} \mbf h \\
		\nabla \mathcal{L}^2_{\mbf f_i, \mbf f_j} \mbf h \\
		\vdots
	\end{bmatrix},
\end{equation}
where $i, j \in [0, 1, \cdots, m]$.
$\mathcal{O}$ can be viewed as a codistribution spanned by row vectors (aka covectors).
That is, $\mathcal{O}$ can be written as 
\begin{equation}
	\mathcal{O} = \mathrm{span}\{\nabla \mathcal{L}^0 \mbf h, \nabla \mathcal{L}^1_{\mbf f_i} \mbf h, \nabla \mathcal{L}^2_{\mbf f_i, \mbf f_j} \mbf h, \cdots\}
\end{equation}

The observability property of the affine-input system can be revealed by incrementally building up the codistribution and checking its dimension.
Denote the codistribution with gradients of Lie derivatives up to order $k$ by 
$\mathcal{O}_k$, \eg, $\mathcal{O}_0 = \mathrm{span}\{ \nabla \mbf h \}$ and 
$\mathcal{O}_1 = \mathrm{span}\{\nabla \mbf h,
\nabla \mathcal{L}^1_{\mbf f_0} \mbf h, \nabla \mathcal{L}^1_{\mbf f_1} \mbf h, \cdots, \nabla \mathcal{L}^1_{\mbf f_m} \mbf h\}$.
As deduced in \cite[Algorithm 4.2]{martinelliObservability2020}, if $\mathrm{rank}(\mathcal{O}_{k-1}) = \mathrm{rank}(\mathcal{O}_k)$, the incremental procedure completes.
By then, if $\mathrm{rank}(\mathcal{O}_{k}) = \dim(\mbf x)$, the system is weakly observable.

To simplify observability analysis, it is beneficial to proactively identify 
observable dimensions of the state $\mbf x$ in building up the codistribution.
Considering the codistribution at step $k$, $\mathcal{O}_k$, this can be done by the following property.
By Lemma 1 in \cite{huaiObservabilityAnalysisKeyframebased2022}, a component of $\mbf x$, $\mbf x_i$, is weakly observable 
if removing the column corresponding to $\mbf x_i$ from $\mathcal{O}_k$ reduces its rank by 1.

In the end, we note that the above analysis requires that the control inputs are unconstrained and that the control inputs and the observations are timestamped by the same clock.
The following will try to address the two limitations.

\section{Observability Analysis with Constraints} \label{sec:woconstraints}
This section presents our procedures to analyze observability when the nonlinear system is subject to linear constraints.
We first define the epitome systems with constraints of increasing complexity which extend the constraint-free standard system \eqref{eq:affine-input-sys}, and then propose the procedures to analyze their observability.
At last, we also list findings about observability of the time offset.

\subsection{Extended Systems}
For convenience, the standard affine-input system is repeated here.
The dynamics of its state $\mbf x$ is affine in the control inputs $\mathcal{U} = \{u_i \vert i=1, 2, \cdots, m\}$. 
With a bit abuse of notation, the vector of control inputs $\mbf u$ is used interchangeably with $\mathcal{U}$.
\subsubsection{Standard System 1}
In the standard system, the observation is a function of just the state vector $\mbf x$.
\begin{equation}
	\label{eq:standard-sys}
		\dot{\mbf{x}} = \mbf{f}_0(\mbf x) + \sum_{i = 1}^{m}\mbf f_i(\mbf x) u_i, \quad
		\mbf{y}(t) = \mbf h(\mbf x(t)).
\end{equation}

\subsubsection{System 2 with a zero constraint on the state}
There is a 1-D constraint $0 = c(\mbf x)$ on the state $\mbf{x}$.
\begin{equation}
	\label{eq:zero-constraint-x-sys}
	\begin{split}
		\dot{\mbf{x}} &= \mbf{f}_0(\mbf x) + \sum_{i = 1}^{m}\mbf f_i(\mbf x) u_i, \\
		\mbf{y}(t) &= \mbf h(\mbf x(t)), \\
		0 &= c(\mbf{x}).
	\end{split}
\end{equation}

\subsubsection{System 3 with a constant constraint on the state}
There is a 1-D constraint $d = c(\mbf x)$ on the state where $d$ is an unknown constant.
\begin{equation}
	\label{eq:constraint-x-sys}
	\begin{split}
		\dot{\mbf{x}} &= \mbf{f}_0(\mbf x) + \sum_{i = 1}^{m}\mbf f_i(\mbf x) u_i, \\
		\mbf{y}(t) &= \mbf h(\mbf x(t)), \\
		d &= c(\mbf{x}).
	\end{split}
\end{equation}

\subsubsection{System 4 with a zero constraint on the state and affine in the inputs}
There is a 1-D constraint $0 = c(\mbf x, \mathcal{U}_c)$ on the state $\mbf{x}$ and a input subset $\mathcal{U}_c \in \mathcal{U}$.
We also assume that the constraint is affine in the control inputs.
\begin{equation}
	\label{eq:zero-constraint-x-u-sys}
	\begin{split}
		\dot{\mbf{x}} &= \mbf{f}_0(\mbf x) + \sum_{i = 1}^{m}\mbf f_i(\mbf x) u_i, \\
		\mbf{y}(t) &= \mbf h(\mbf x(t)), \\
		0 &= c(\mbf x, \mathcal{U}_c) = c_0(\mbf{x}) + \sum_{i=1}^{l} c_i(\mbf x) u_l \quad u_l \in \mathcal{U}_c
	\end{split}
\end{equation}

\subsubsection{System 5 with a constant constraint on the state and affine in the inputs}
There is a 1-D constraint $d = c(\mbf x, \mathcal{U}_c)$ on the state $\mbf{x}$ and a control input subset $\mathcal{U}_c\in \mathcal{U}$, 
where $d$ is a unknown constant.
We further assume that the constraint is affine in the control inputs.
\begin{equation}
	\label{eq:constraint-x-u-sys}
	\begin{split}
		\dot{\mbf{x}} &= \mbf{f}_0(\mbf x) + \sum_{i = 1}^{m}\mbf f_i(\mbf x) u_i, \\
		\mbf{y}(t) &= \mbf h(\mbf x(t)), \\
		d &= c_0(\mbf{x}) + \sum_{i=1}^{l} c_i(\mbf x) u_l \quad u_l \in \mathcal{U}_c
	\end{split}
\end{equation}

\subsection{Conversion to the Standard Form}
\label{subsec:conversion}
For each of the five systems, our goal is to
check if the state $\mbf x(t)$ is weakly observable given control inputs
$\mathbf{u}$ and observations in time interval $[t, t + \Delta t]$.
Since we already know how to check the local observability for System 1 \eqref{eq:standard-sys}, the idea is to 
convert a system with linear constraints into the standard form.
The guideline for the conversion is to keep the information of the system before and after the conversion the same.

\subsubsection{Conversion of System 2}
We start with System 2 \eqref{eq:zero-constraint-x-sys}.
One may be tempted to simply add the constraint $0 = c(\mbf x)$ to the set of observations. However, the subsequent observability analysis may consider $c(\mbf x)$ to be time-varying as $\mbf y(t)$ in $\mbf y(t) = \mbf h(\mbf x(t))$, effectively undoing the constraint.
Therefore, we choose a $x_p \in \mbf x$, express $x_p$ in terms of the other variables involved in $c(\mbf x)$ as $x_p = c'(\mbf x_r)$, and finally substitute $c'$ for $x_p$ in the system equations, \ie,

\begin{equation}
	\label{eq:zero-constraint-x-sys-converted}
	\begin{split}
		\dot{\mbf{x}'} &= \mbf{f}_0'(\mbf x') + \sum_{i = 1}^{m}\mbf f_i'(\mbf x') u_i, \\
		\mbf{y}(t) &= \mbf h'(\mbf x'(t)),
	\end{split}
\end{equation}
where $\mbf x' = \mbf x \backslash x_p$.

\subsubsection{Conversion of System 3}
System 3 can be converted to the standard form in two steps, first to System 2, and then to System 1.
The first step is done by adding $d$ as an unknown parameter to the state, and the system becomes 
\begin{equation}
	\label{eq:constraint-x-sys-converted}
	\begin{split}
		\begin{bmatrix}
			\dot{\mbf{x}} \\
			\dot{d}
		\end{bmatrix}
		&= 
		\begin{bmatrix}
			\mbf{f}_0(\mbf x) + \sum_{i = 1}^{m}\mbf f_i(\mbf x) u_i \\
			0
		\end{bmatrix}, \\
		\begin{bmatrix}
			\mbf{y}(t) \\
			0
		\end{bmatrix}
		&= \begin{bmatrix}
			\mbf h(\mbf x(t)) \\
			c(\mbf{x}) - d
		\end{bmatrix}.
	\end{split}
\end{equation}
The second step is the same as for System 2 except that the chosen $x_p$ should not be $d$.

\subsubsection{Conversion of System 4}
\label{subsubsec:conv-sys4}
For System 4, to maintain the affine-input property, instead of replacing $x_p$, we pick up a control input $u_o \in \mathcal{U}_c = \{u_o, \mbf u_r\}$
(subscript ${\cdot}_o$ for removal) and remove it from the differential equation for $\mbf x$ with the steps:
\begin{enumerate}
	\item choose one $u_o \in \mathcal{U}_c$, and express it by rearranging the constraint,  $u_o = c^\prime(\mbf x, \mathbf{u}_r)$ (note that $c'$ is affine in $\mathbf{u}_r$);
	\item substitute the expression $c^\prime$ for $u_o$ in the process equation $\dot{\mbf x}$.
\end{enumerate}

Next, since $u_o$ is known, we will take $u_o = c^\prime(x, \mathbf{u}_r)$ as an observation. However, unlike observations in the standard form, it has control inputs $\mathbf{u}_r$. Therefore, we will factor these control inputs out of this observation with the steps:
\begin{enumerate}
	\setcounter{enumi}{2}
	\item treat control inputs $\mathbf{u}_r$ as state variables and add to the state vector $\mbf x$;
	
	\item remove $\mathcal{U}_c$ from the control inputs, 
	add time derivatives of $\mathbf{u}_r$ to the control inputs,
	
	\item finally, add $u_o$ and $\mathbf{u}_r$ as observations to the system.
\end{enumerate}

In the end, the resulting system has a state vector
$\mathbf{x}^\prime = [\mbf x; \mathbf{u}_r]$,
and control inputs 
$\{\mathcal{U} \backslash \mathcal{U}_c, \dot{\mathbf{u}}_r\}$.
The new system can be written as
\begin{equation}
	\label{eq:zero-constraint-x-u-sys-converted}
	\begin{split}
		\frac{d}{dt}\begin{bmatrix}
			\mbf{x} \\ 
			\mbf{u}_r
		\end{bmatrix}
		&= \begin{bmatrix}
			\mbf{f}_0^\prime(\mbf x) + \sum_{i \neq o}^{m} \mbf f_i^\prime(\mbf x) u_i, \\
			\dot{\mbf{u}}_r
		\end{bmatrix} \\
		\begin{bmatrix}
			\mbf{y}(t) \\ 
			\mbf{u}_r \\
			u_o
		\end{bmatrix}
		&= \begin{bmatrix}
			\mbf h(\mbf x(t)) \\
			\mbf{u}_r \\
			c^\prime(x, \mbf u_r)
		\end{bmatrix}
	\end{split}
\end{equation}
where $\mbf f_0^\prime$ and $\mbf f_i^\prime$ correspond to $\mbf f_0$ and $\mbf f_i$ after $c^\prime(\mbf x, \mbf{u}_r)$ is plugged in, respectively.
It is easy to confirm that the above system is in the standard form \eqref{eq:standard-sys}.

\subsubsection{Conversion of System 5}
\label{subsubsec:conv-sys5}
The system 5 can be converted to the standard form in two steps.
First, add $d$ as a unknown parameter to the state, leading to
\begin{equation}
	\label{eq:constraint-x-u-sys-converted}
	\begin{split}
		\begin{bmatrix}
			\dot{\mbf{x}} \\
			\dot{d}
		\end{bmatrix}&= 
		\begin{bmatrix}
			\mbf{f}_0(\mbf x) + \sum_{i = 1}^{m}\mbf f_i(\mbf x) u_i \\
			0
		\end{bmatrix}, \\
		\begin{bmatrix}
			\mbf{y}(t) \\ 0
		\end{bmatrix} &= \begin{bmatrix}
			\mbf h(\mbf x(t)), \\
			c_0(\mbf{x}) + \sum_{i=1}^{l} c_i(\mbf x) u_l - d
		\end{bmatrix} \quad u_l \in \mathcal{U}_c.
	\end{split}
\end{equation}
Second, convert the new system to the standard form as for System 4.
 
The above conversion procedures can be applied to multi-dimensional constraints by dealing with each dimension sequentially.

\subsection{Observability of Time Offset}
\label{subsec:obs-time-offset}
The observability of the time offset between control inputs and output observations for a nonlinear system affine in inputs has been conducted in \cite{huaiObservabilityAnalysisKeyframebased2022} and the conclusion is repeated below.
The nonlinear system with time offset can be expressed by 
\begin{equation}
	\label{eq:unsync-sys}
		\dot{\mbf{x}} = \mbf{f}_0(\mbf x) + \sum_{i = 1}^{m}\mbf f_i(\mbf x) u_i, \quad
		\mbf{y}(t) = \mbf h(\mbf x(t + t_d)),
\end{equation}
where $t_d$ is the time offset.
Given control inputs
$\mathbf{u}$ in time interval $[t + t_d, t + \Delta t + t_d]$ and observations in $[t, t+\Delta t]$, the time offset is weakly unobservable when the system goes through constant control inputs or constant observations.

\section{Application to Visual Inertial Odometry under Constraints}
To validate the proposed procedures, we consider a VIO system that fuses camera observations of opportunistic landmarks and measurements of an IMU (consisting of a 3-axis accelerometer and a 3-axis gyroscope) to
track motion of the camera-IMU platform.

\textbf{Coordinate Frames}: The world frame $\{W\}$ is chosen at the start of the VIO with z-axis along the negative gravity, such that $\mbf g^W = [0, 0, -g]$ where $g$ is the gravity magnitude.
We assume that the IMU is well calibrated but subject to constant biases.
The IMU frame $\{S\}$ is realized by the accelerometer triad.
The camera frame $\{C\}$ is at the camera optical center.
The body frame $\{B\}$ is set to be $\{S\}$.

\textbf{Rotation Parameterization}: The Hamilton quaternion as in Eigen library and ROS \cite{solaQuaternion2017} is used.

A quaternion $\mbf q_{AB} = [q_w, q_x, q_y, q_z]\tran$ can be converted to the rotation matrix
$\mbf R_{AB}$ (orientation of frame $\{B\}$ in frame $\{A\}$) by
\begin{equation}
		\mbf R_{AB}(\mbf q_{AB}) = \begin{bmatrix}
			q_w^2 + q_x^2 - q_y^2 - q_z^2 & 2(q_x q_y - q_w q_z) & 2 (q_x q_z + q_w q_y) \\
			2 (q_x q_y + q_w q_z) & q_w^2 - q_x^2 + q_y^2 - q_z^2 & 2(q_y q_z - q_w q_x) \\
			2(q_x q_z - q_w q_y) & 2 (q_y q_z + q_w q_x) & q_w^2 - q_x^2 - q_y^2 + q_z^2
		\end{bmatrix}.
\end{equation}
For easy result interpretation, the extra freedom of a quaternion is trimmed by expressing it by
$\bar{\mbf q} = [q_x, q_y, q_z]\tran$ whose correponding unit quaternion is $\mbf q = [1, q_x, q_y, q_z]\tran / \sqrt{[1, q_x, q_y, q_z]\cdot [1, q_x, q_y, q_z]\tran}$.
The concern of singularity of $\bar{\mbf q}$ is appeased by considering that the symbolic analysis procedure assumes general values for variables.
Otherwise, 4-parameter quaternions can be used without affecting the conclusions.

\textbf{State Vector}: For the VIO system, the state vector consists of velocity of $\{B\}$ relative to $\{W\}$ expressed in $\{B\}$, $\mbf v^B$,
the gravity vector in $\{B\}$, $\mbf g^B$, 
IMU biases $\mbf b = [\mbf b_g\tran, \mbf b_a\tran]\tran$, 
the relative pose of the IMU in the camera frame, $[\mbf p_{CB}, \bar{\mbf q}_{CB}]$,
and parameters of a landmark $L$ with coordinates $\mbf p_L^C = [p_x^C, p_y^C, p_z^C]\tran$, $\mbs{\gamma}$ and $\rho$,
\begin{equation}
	\mbf x = [\mbf v^{B}, \mbf g^B, \mbf b, \mbf p_{CB}, \bar{\mbf q}_{CB}, \mbs{\gamma}, \rho],
\end{equation}
where  $\mbs{\gamma} \triangleq [p_x^C / p_z^C, p_y^C / p_z^C]$ and
$\rho \triangleq 1 / p_z^C$.
The state variables are chosen so that the unobservable dimensions of a general VIO system (global translation and the rotation angle about gravity) are not included.
The gravity vector $\mbf g^B$ encodes the roll and pitch angles of the system and the gravity magnitude.
Without loss of generality, only one landmark is chosen for the analysis.

\textbf{Continuous System Model}: For the sake of observability analysis, all the noises in the process and observation equations are omitted.
The perfect IMU measurements $\mbf a_t$ and $\mbs \omega_t$ can be obtained from the 
IMU readings $\mbf a_m$ and $\mbs \omega_m$ by
\begin{equation}
		\mbf a_t = \mbf a_m - \mbf b_a \quad 
		\mbs \omega_t = \mbs \omega_m - \mbs b_g
\end{equation}
By basic derivation, the differential equation of the VIO system is found as
\begin{equation}
\label{eq:vio-process}
	\begin{bmatrix}
		\dot{\mbf v}^{B} \\ \dot{\mbf g}^{B} \\ \dot{\mbf b} \\ \dot{\mbf p}_{CB} 
		\\ \dot{\bar{\mbf q}}_{CB} \\ \dot{\mbs{\gamma}} \\ \dot{\rho}
	\end{bmatrix} =
	\begin{bmatrix}
		\mbf{v}^B \times \mbs{\omega}_t + \mbf{a}_t + \mbf{g}^B \\
		\mbf g^B \times \mbs{\omega}_t \\
		\mbf 0_6 \\ \mbf 0_3 \\ \mbf 0_3 \\
		\begin{array}{c}
			C_{\mbs{\gamma}\rho}[
			(\mbf{R}_{CB} \mbs{\omega}_t) \times (\rho \mbf{p}_{CB} - \bar{\mbs{\gamma}}) -\\
			\mbf{R}_{CB} \rho \mbf{v}^B]
		\end{array}
	\end{bmatrix}
\end{equation}
where the shorthand symbols are 
\begin{equation}
	\bar{\mbs{\gamma}} = [\mbs{\gamma}\tran \quad 1]\tran, \quad
	C_{\mbs{\gamma} \rho} = \begin{bmatrix}
		\mbf{I}_2 & -\mbs{\gamma} \\ 0_{1\times2} & -\rho
	\end{bmatrix}.
\end{equation}
The control inputs to the system are $\mbf u = \{\mbs \omega_m, \mbf a_m\}$.
The observations of the system are the camera reprojections and the gravity norm.
We assume that the camera intrinsic parameters are known, thus we can undistort and normalize the image observation at image plane $z=1$, giving $\mbf h_1 = \mbs \gamma$. In summary, the observations are
\begin{equation}
\mbf{h}_1 = \mbs \gamma \quad
h_2 = {\mbf {g}^B}\tran \mbf g^B.
\label{eq:vio-observations}
\end{equation}

\subsection{Observability under Degenerate Motion}
In the following, we study the VIO system \eqref{eq:vio-process}-\eqref{eq:vio-observations} under three types of constraints corresponding to three degenerate motion types, and identify the corresponding unobservable state variables by using the method in Section \ref{subsec:conversion}.
The three degenerate motion types are constant local linear acceleration,
single-axis rotation (and otherwise general translation), and pure translation.

The constant local acceleration involves a 3-D constraint given by
\begin{equation}
\label{eq:const-local-accel}
\mathrm{const} = \mbf a^B = \mbf a_t + \mbf g^B = \mbf a_m - \mbf b_a + \mbf g^B.
\end{equation}

When the platform is constrained to go through single-axis rotation (\eg, general motion on a 2-D planar), the constant global rotation axis constraint is expressed by
$\mathrm{const} = \omega^W_{WB}/\sqrt{(\omega^W_{WB})\tran \omega^W_{WB}}$.
Without loss of generality, we assume that the rotation axis is aligned with the z-axis of the IMU frame, thus, the constraint becomes
\begin{equation}
\label{eq:single-axis}
   [0, 0]\tran  = [\omega_{t x}, \omega_{t y}]\tran = (\mbs \omega_m - \mbf b_g)_{1:2},
\end{equation}
where $\mbs \omega_t = [\omega_{tx}, \omega_{ty}, \omega_{tz}]\tran$.
When the rotation axis is not along the IMU z-axis, we can define a $\{B'\}$ frame at the same origin as $\{B\}$ and with z-axis along the rotation axis.
Denoting the rotation between the $\{B'\}$ and $\{B\}$ by $\mbf R_{BB'}$, we see that the single-axis rotation constraint relates to the gyroscope reading by
\begin{equation}
\mbs \omega_m = \mbf R_{BB'} [0, 0, \lambda]\tran + \mbf b_g,
\end{equation}
where $\lambda$ is the time-varying angular rate.
The corresponding constraints after rearranging the terms are
\begin{equation}
\label{eq:single-axis-ext}
   [0, 0]\tran = [\mbf R_{BB'}\tran(\mbs \omega_m - \mbf b_g)]_{1:2}.
\end{equation}

The pure translation motion means that the angular rate is zero,
\begin{equation}
\label{eq:pure-translation}
\mbf 0_{3\times1} =\mbs \omega_t = \mbs \omega_m - \mbf b_g.
\end{equation}

Using the procedures presented in Section \ref{subsec:conversion}, we incorporate these three types of constraints separately into the unconstrained VIO system \eqref{eq:vio-process}-\eqref{eq:vio-observations} and convert each of the three specific systems into the standard form.
For the motion of constant local acceleration \eqref{eq:const-local-accel}, we use the procedure for System 5.
For the single-axis rotation constraint \eqref{eq:single-axis} and the pure translation constraint \eqref{eq:pure-translation}, the procedure for System 4 is applied.

Then, each converted system is automatically analyzed by an observability checking procedure similar to Algorithm 4.2 \cite{martinelliObservability2020} which was implemented in MATLAB using three key symbolic functions $\mathrm{diff}()$, $\mathrm{rank}()$, $\mathrm{null}()$.
The improvement relative to Algorithm 4.2~\cite{martinelliObservability2020} includes keeping only row vectors spanning the row space of the observability matrix $\mathcal{O}_k$ after each iteration $k$ to restrain the problem size and speed up computation of the null space.

The automatic procedure concludes that the system in \eqref{eq:vio-process}-\eqref{eq:vio-observations} is fully observable, when the platform goes through general motion, as expected.
With the constant local acceleration, the metric scale of the system becomes additionally unobservable, confirming statements in \cite{wuVINS2017} and \cite{yangDegenerateMotionAnalysis2019}.
The automatic procedure identifies the corresponding 1-column null space $\mbf N$,
\begin{equation}
\begin{bmatrix}
(\mbf v^B)\tran & (\mbf g^B)\tran & \mbf 0_3\tran & -(\mbf a_t + \mbf g^B)\tran & (\mbf p_{CB})\tran & \mbf 0_{1\times5} & -\rho
\end{bmatrix}\tran.
\end{equation}

For the single-axis rotation, our procedure found that the z-component of $\mbf p_{CB}$ is additionally unobservable as concluded in \cite{wuVINS2017} and \cite{yangDegenerateMotionAnalysis2019}, with a 1-column null space,
\begin{equation}
\begin{bmatrix}
\mbf 0_{1\times12} \quad 2(q_y + q_x q_z) \quad -2(q_x - q_y q_z) \quad - (q_x^2 + q_y^2 - q_z^2 - 1) \quad \mbf 0_{1\times 6}
\end{bmatrix}\tran,\end{equation}
where $\bar{\mbf q}_{CB} = [q_x, q_y, q_z]$.
Since a perturbation $\epsilon \xi$ in the null space on $\mbf x$, \ie, $\mbf x^\prime = \mbf x + \epsilon \xi, \xi \in \mbf N$, does not affect the system, in this case, we see that the only vector in the null space changes $\mbf p_{CB}$ by the amount $\epsilon \mbf R_{CB} [0, 0, 1]\tran$ which corresponds to an increment on the z-component of $\mbf p_{CB}$.

For the pure translation, our procedure found that $\mbf p_{CB}$, $\mbf g^B$, and $\mbf b_a$ are indeterminable (\ie, unable to be verified as observable by Lemma 1 in \cite{huaiObservabilityAnalysisKeyframebased2022}).
There is a 5-column null space given by
\begin{equation}\mbf N\tran =
\left[
\begin{array}{c|ccc|c|ccc|cc}
\mbf 0_{3\times 3} & \mbf 0_{3\times 1} & \mbf 0_{3\times 1} & \mbf 0_{3\times 1} & \mbf 0_{3\times 3} & \mbf 0_{3\times 1} & \mbf 0_{3\times 1} & \mbf 0_{3\times 1} & \mbf I_3 & \mbf 0_{3\times 6} \\
\mbf 0_{1\times 3} & -g^B_y & g^B_x & 0 & \mbf 0_{1\times 3} & -g^B_y & g^B_x & 0 & 0_{1\times 3} & 0_{1\times 6} \\
\mbf 0_{1\times 3} & -g^B_z & 0 & g^B_x & \mbf 0_{1\times 3} & -g^B_z & 0 & g^B_x & 0_{1\times 3} & 0_{1\times 6}
\end{array}\right].
\end{equation}
It is easy to confirm that the first 3 rows of $\mbf N\tran$ correspond to $\mbf p_{CB}$, and the last two rows correspond to roll and pitch of the platform, meaning that $\mbf p_{CB}$ and $\mbf q_{WB}$ are unobservable.
Because of the correlation between $\mbf g^B$ and $\mbf b_a$, both $\mbf g^B$ and $\mbf b_a$ are indeterminable.
In summary, the findings by our analysis procedure and by using the linearized model are listed in Table~\ref{tab:degenerate-findings}.

\begin{table}[!htb]
\centering
\caption{Unobservable state variables of a visual inertial odometry system with extrinsic calibration under three degenerate motion types, obtained by the observability matrix built from the linearized model and by that built from Lie derivatives of the nonlinear system. col($\mbf N$) denotes the number of columns of the null space matrix $\mbf N$.}
\label{tab:degenerate-findings}
\begin{tabular}{@{}ll|l|l|l@{}}
\toprule
\multicolumn{2}{c|}{\textbf{\begin{tabular}[c]{@{}l@{}}Unobservable \\ state variables\end{tabular}}}           & \textbf{\begin{tabular}[c]{@{}l@{}}Rotation about\\z-axis of $\{B\}$\end{tabular}} & \textbf{\begin{tabular}[c]{@{}l@{}}Const. local\\ acceleration\end{tabular}}                                                 & \textbf{\begin{tabular}[c]{@{}l@{}}No rotation\end{tabular}}                                                       \\ \midrule
\multicolumn{2}{c|}{\begin{tabular}[c]{@{}l@{}}Linearized model \\incl. $\mbf p_{WB}$ and yaw\\ \cite{wuVINS2017} \cite{yangDegenerateMotionAnalysis2019} \end{tabular}} & \begin{tabular}[c]{@{}l@{}}$\mathbf{p}_{CB}$ along\\ z-axis\end{tabular}   & scale & \begin{tabular}[c]{@{}l@{}}global rotation,\\ $\mathbf{p}_{CB}$\end{tabular}                                             \\ \midrule
\multicolumn{1}{c|}{\multirow{2}{*}[-1em]{\begin{tabular}[c]{@{}l@{}}Nonlinear model\\ (our result)\end{tabular}}}
& Indeterminable & $\mathbf{p}_{CB}$ & $\mbf v^B, \mathbf{b}_a, \mathbf{p}_{CB}, \rho$  & $\mbf g^B, \mathbf{b}_a, \mathbf{p}_{CB}$
\\ \cmidrule(l){2-5}
\multicolumn{1}{c|}{} & Null space & \begin{tabular}[c]{@{}l@{}}col($\mbf N$)=1\end{tabular}                     & \begin{tabular}[c]{@{}l@{}}col($\mbf N$)=1\end{tabular} & \begin{tabular}[c]{@{}l@{}}col($\mbf N$)=5\end{tabular}
\\ \cmidrule(l){2-5}
\multicolumn{1}{c|}{} & \begin{tabular}[c]{@{}l@{}}Unobservable \end{tabular} & \begin{tabular}[c]{@{}l@{}}$\mathbf{p}_{CB}$ along \\ z-axis\end{tabular}                     & \begin{tabular}[c]{@{}l@{}}scale\end{tabular} & \begin{tabular}[c]{@{}l@{}}roll and pitch,\\ $\mathbf{p}_{CB}$\end{tabular} \\ \bottomrule
\end{tabular}
\end{table}

Considering the time offset parameter, we also list in Table~\ref{tab:time-offset-obs-cond} the sufficient conditions when it is unobservable as stated in Section~\ref{subsec:obs-time-offset} and those reported by the related work.
The conditions reported by \cite{yangDegenerateMotionAnalysis2019} is a subset of those identified by our analysis, because the VIO system can be rewritten in terms of alternative sets of control inputs including $[\mbs \omega^B, \mbf a^W]$, and $[\mbs \omega^B, \mbf a^B]$, where $\mbf a^W$ and $\mbf a^B$ are the platform acceleration caused by specific forces expressed in $\{W\}$ and $\{B\}$, respectively.
\begin{table}[!htb]
\centering
\caption{Sufficient conditions when the camera's time offset to the IMU is unobservable, obtained by the linearized model and by the continuous model. $\mbf a^W$ is the platform acceleration due to specific forces in the $\{W\}$ frame.}
\label{tab:time-offset-obs-cond}
\begin{tabular}{@{}l|l@{}}
\toprule
\textbf{\begin{tabular}[c]{@{}l@{}}Time offset unobservable\end{tabular}}                                                                            & \textbf{Sufficient condition}                           \\ \midrule
\multirow{2}{*}
{\begin{tabular}[c]{@{}l@{}}Linearized model \cite{yangDegenerateMotionAnalysis2019}, \cite{yangOnlineSelfcalibrationVisualinertial2022}\end{tabular}} & const. $\boldsymbol \omega^B$ and const. $\mathbf{v}^B$ \\ \cmidrule(l){2-2} 
                                                                                                             & const. $\boldsymbol \omega^B$ and const. $\mathbf{a}^W$ \\ \midrule
\begin{tabular}[c]{@{}l@{}}Nonlinear model (our result)\end{tabular}                                       & const. control inputs \\ \bottomrule
\end{tabular}
\end{table}

\section{Experimental Results}
This section presents simulation results to validate the preceding observability properties.
To easily reproduce the results, the simulation and VIO estimation was carried out in OpenVINS. In simulation, a monocular camera-IMU system moves on a specified trajectory.
Random point landmarks are put in the scene to ensure the camera observes enough features in every view sampled at 10 Hz.
The IMU data are sampled at 400 Hz.

\subsection{Observability with Constrained Motion}
\begin{figure}[htb]
\includegraphics[width=.32\textwidth]{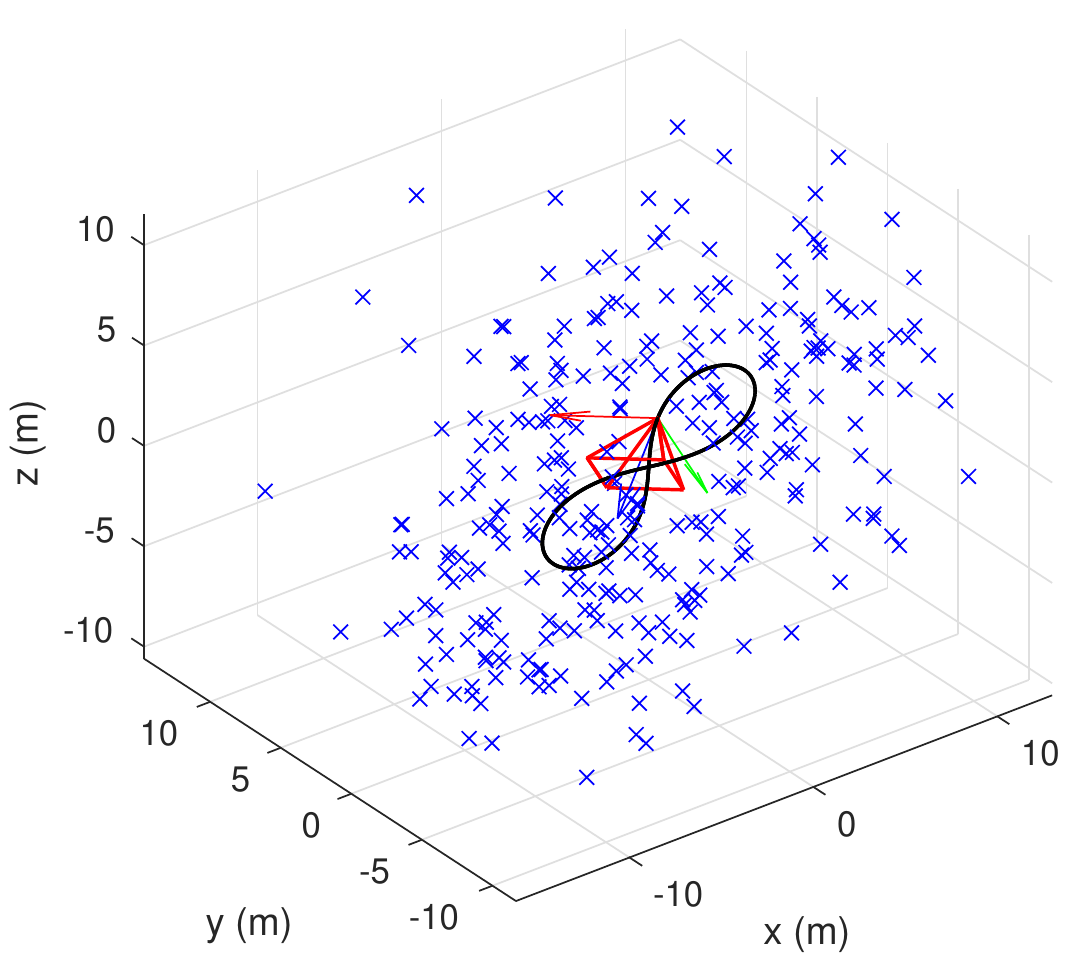}
\includegraphics[width=.32\textwidth]{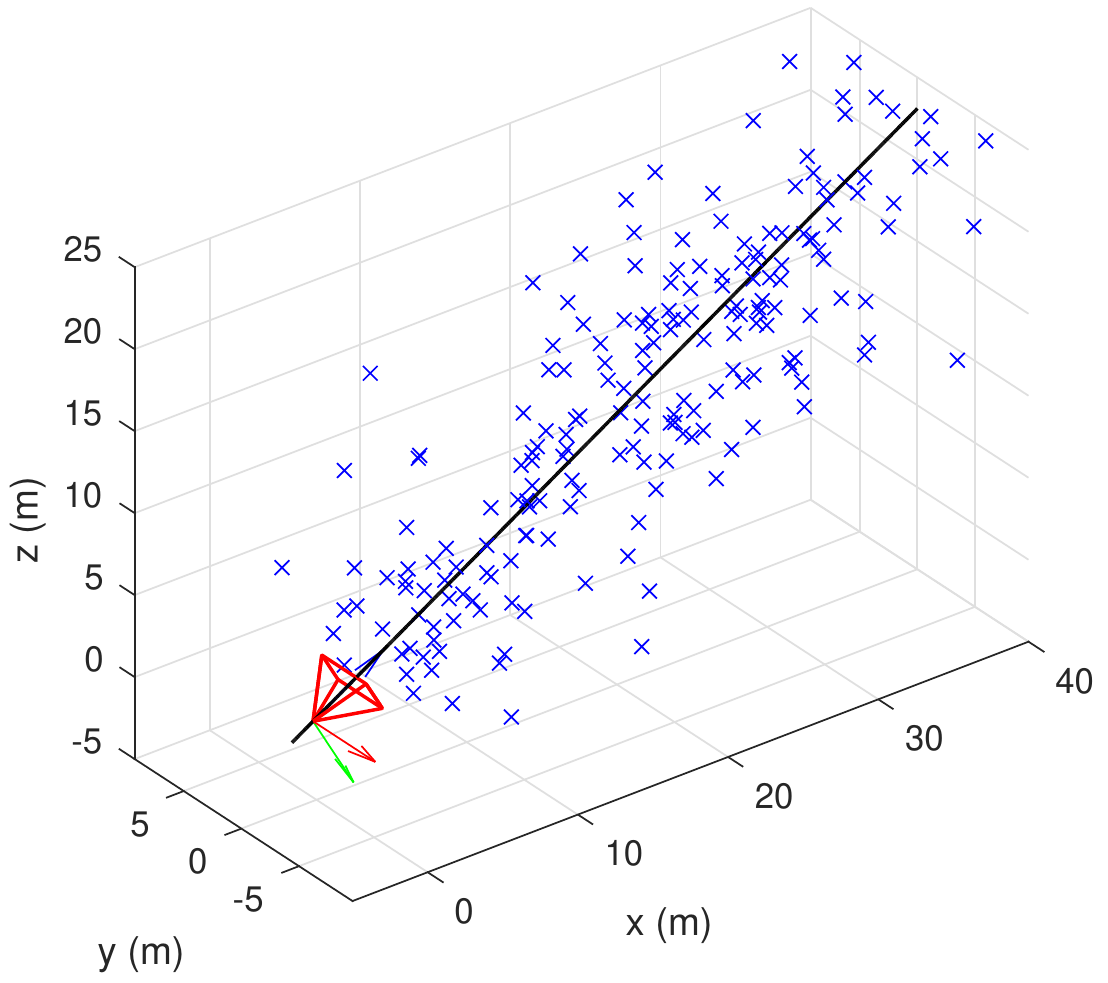}
\includegraphics[width=.32\textwidth]{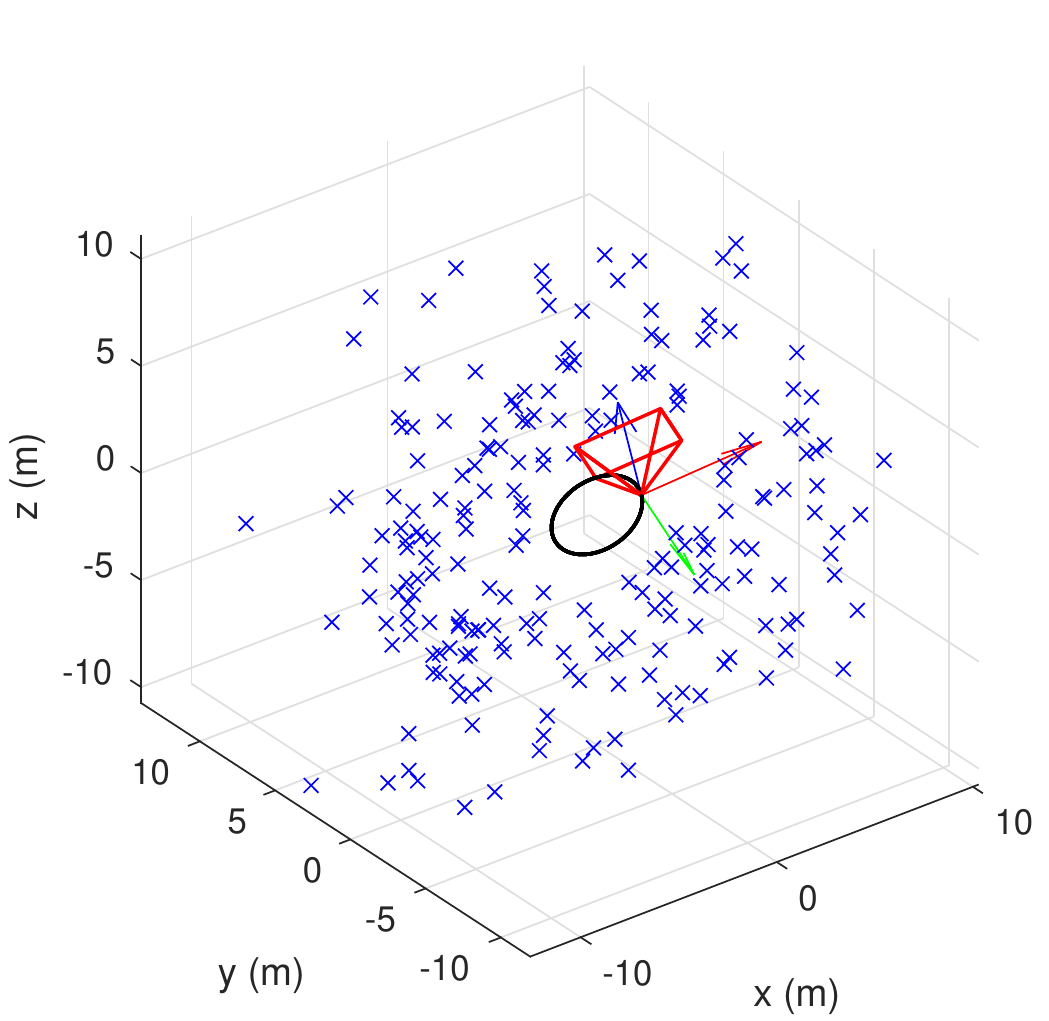}
\centering
\caption{Three planar trajectories (black) on a 30$^\circ$ slope, lemniscate, line segment, and circle, and the random landmarks (blue crosses). The gravity is along the negative z-axis of the $\{W\}$ frame.}
\label{fig:planar-motion}
\end{figure}

To validate the unobservable parameters listed in Table~\ref{tab:degenerate-findings},
we simulate a camera-IMU system moving on a 30$^\circ$ slope in three trajectories, straight line of varying linear accelerations, lemniscate of varying angular rates, and circle of a constant angular rate, as depicted in Fig.~\ref{fig:planar-motion}.
The z-axis of the IMU frame is along the normal of the slope.
The gravity is along the negative z-axis of the $\{W\}$ frame.

\begin{figure}[htb]
\includegraphics[width=.32\textwidth]{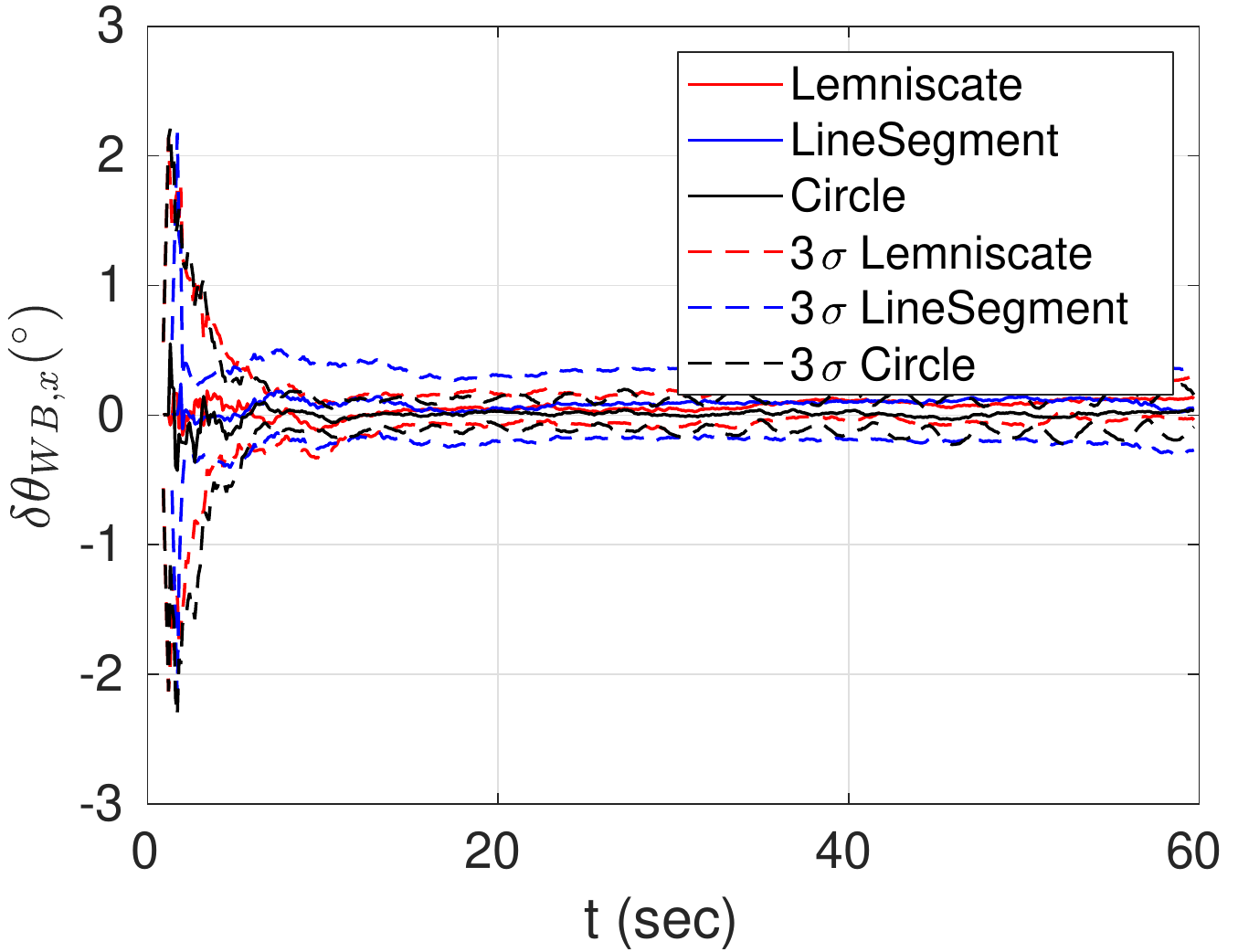}
\includegraphics[width=.32\textwidth]{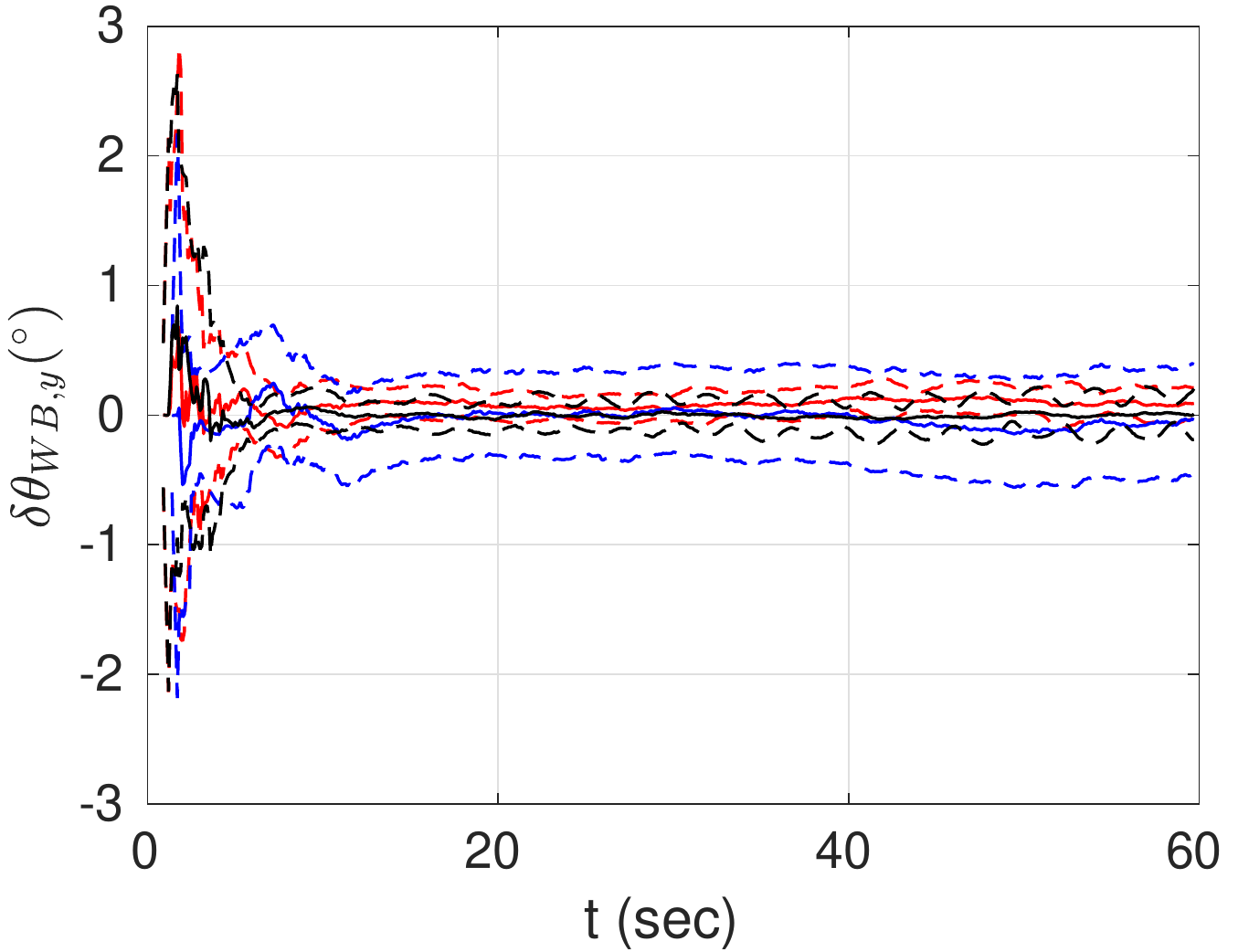}
\includegraphics[width=.32\textwidth]{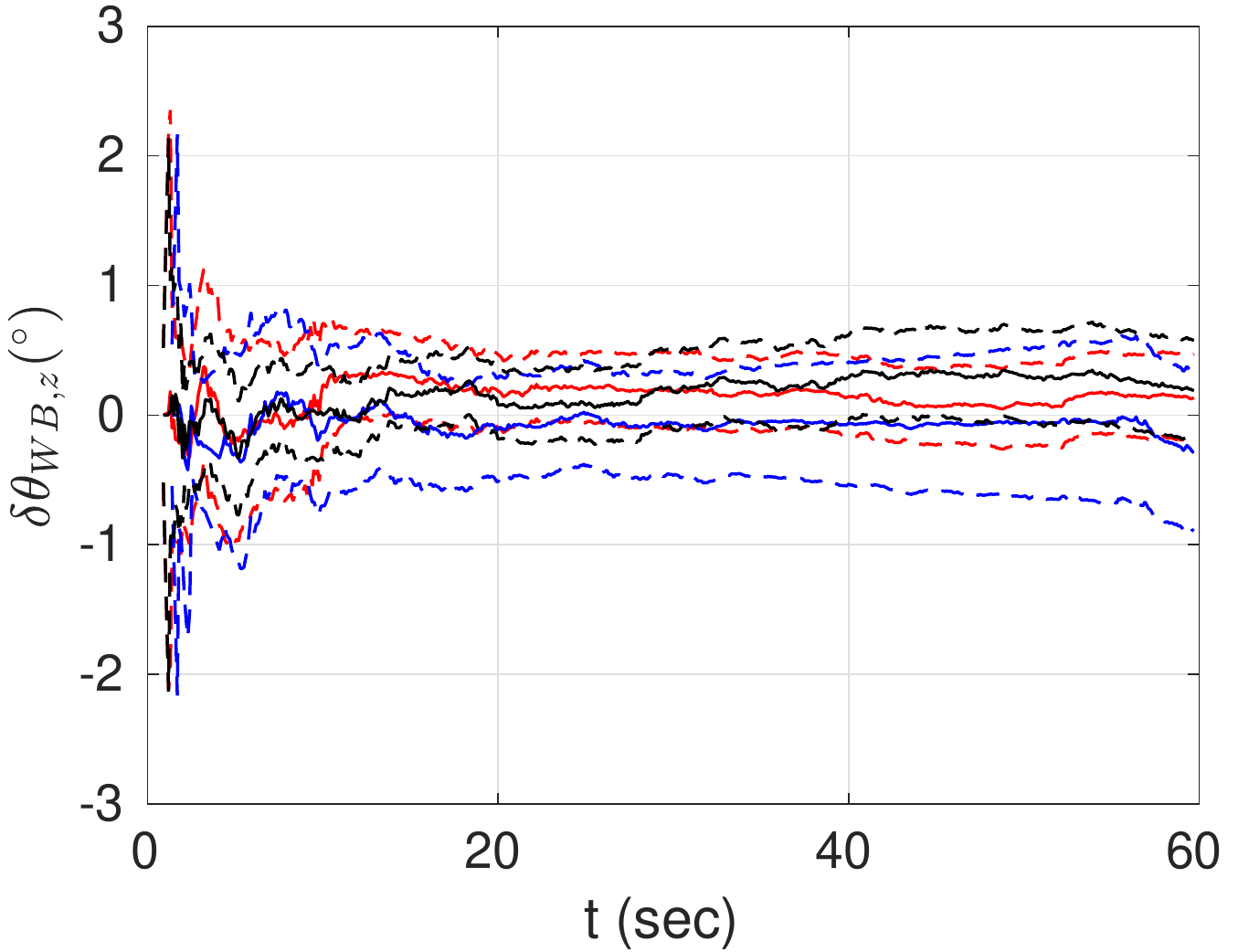}
\includegraphics[width=.32\textwidth]{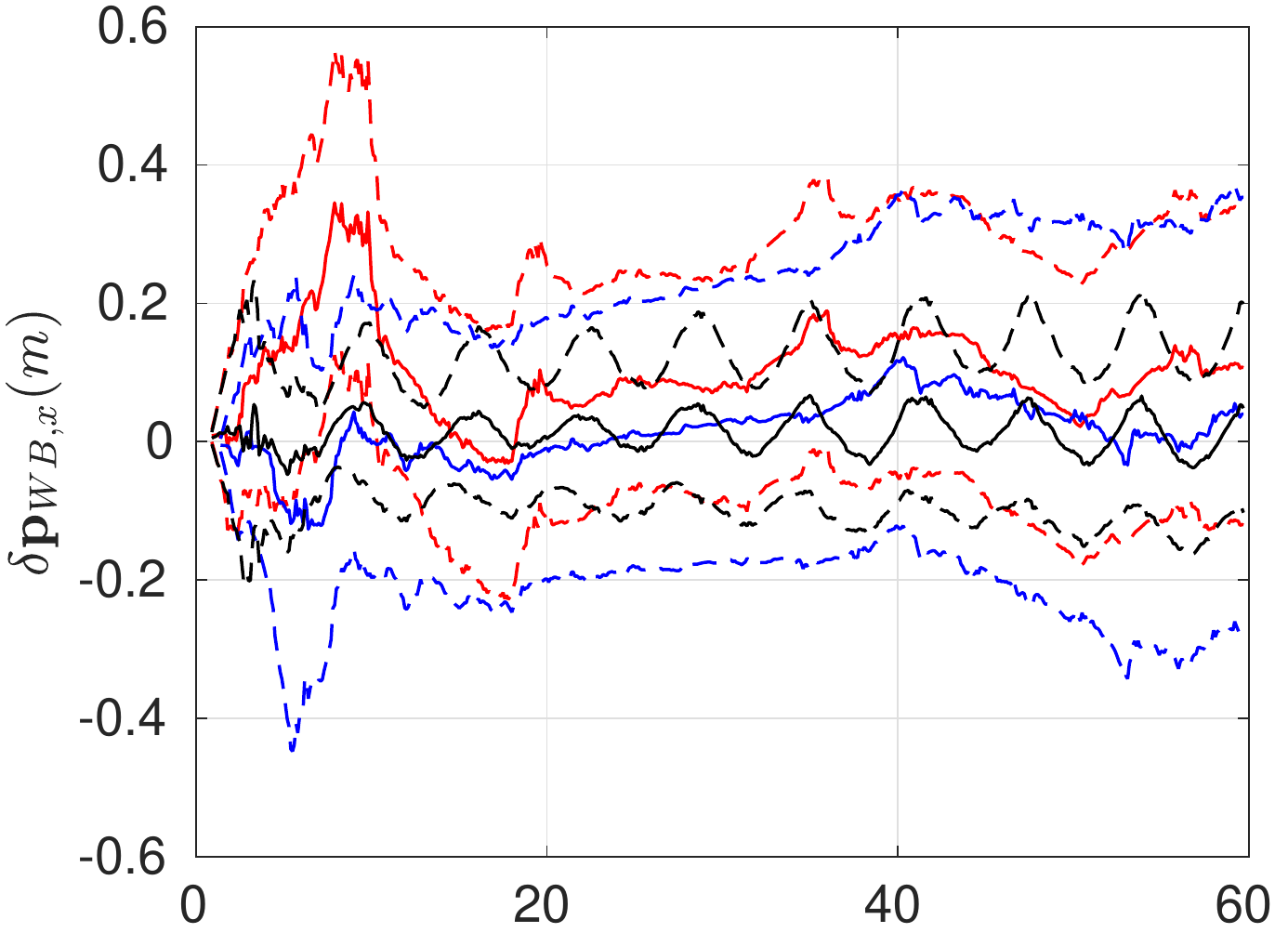}
\includegraphics[width=.32\textwidth]{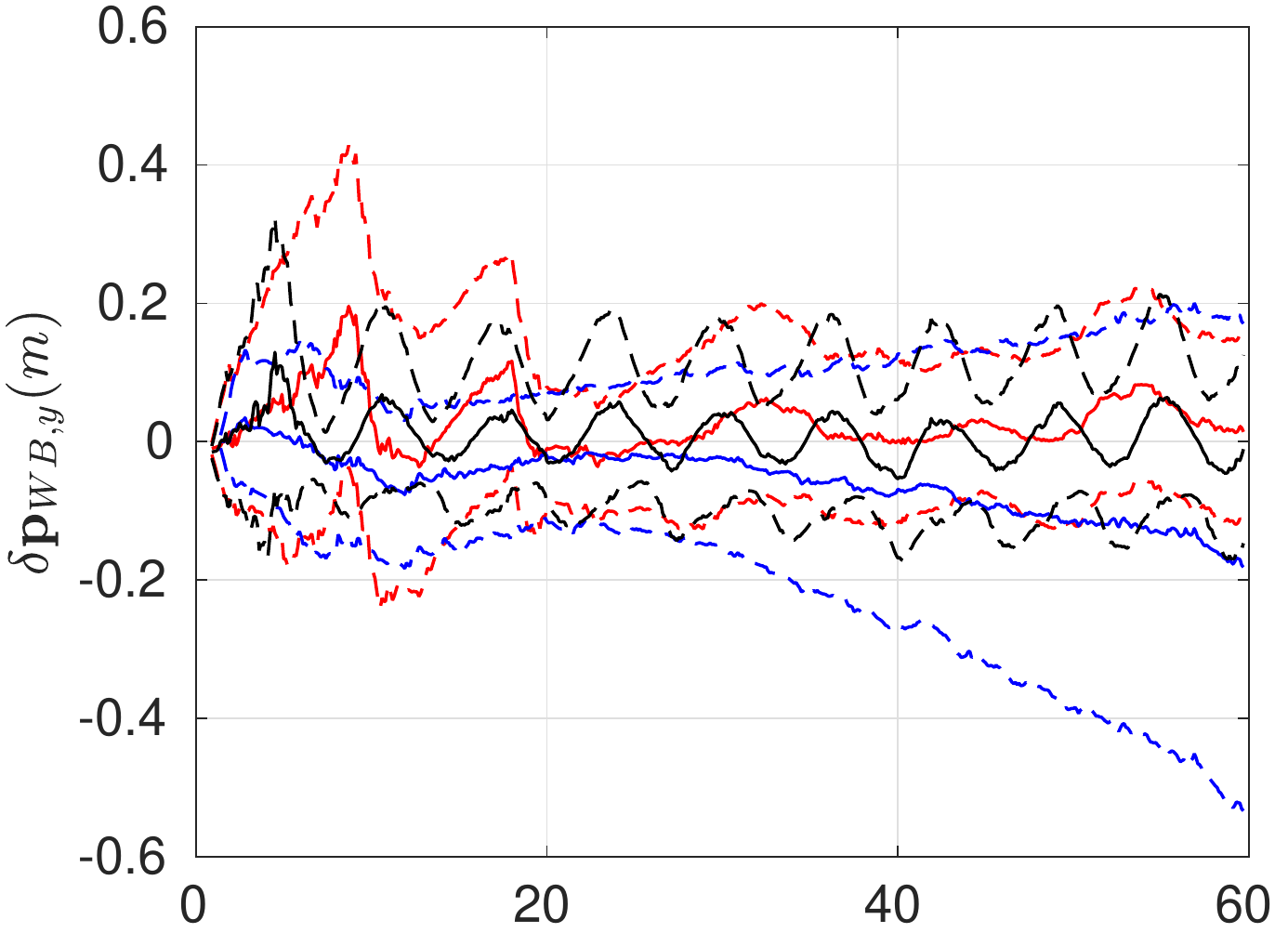}
\includegraphics[width=.32\textwidth]{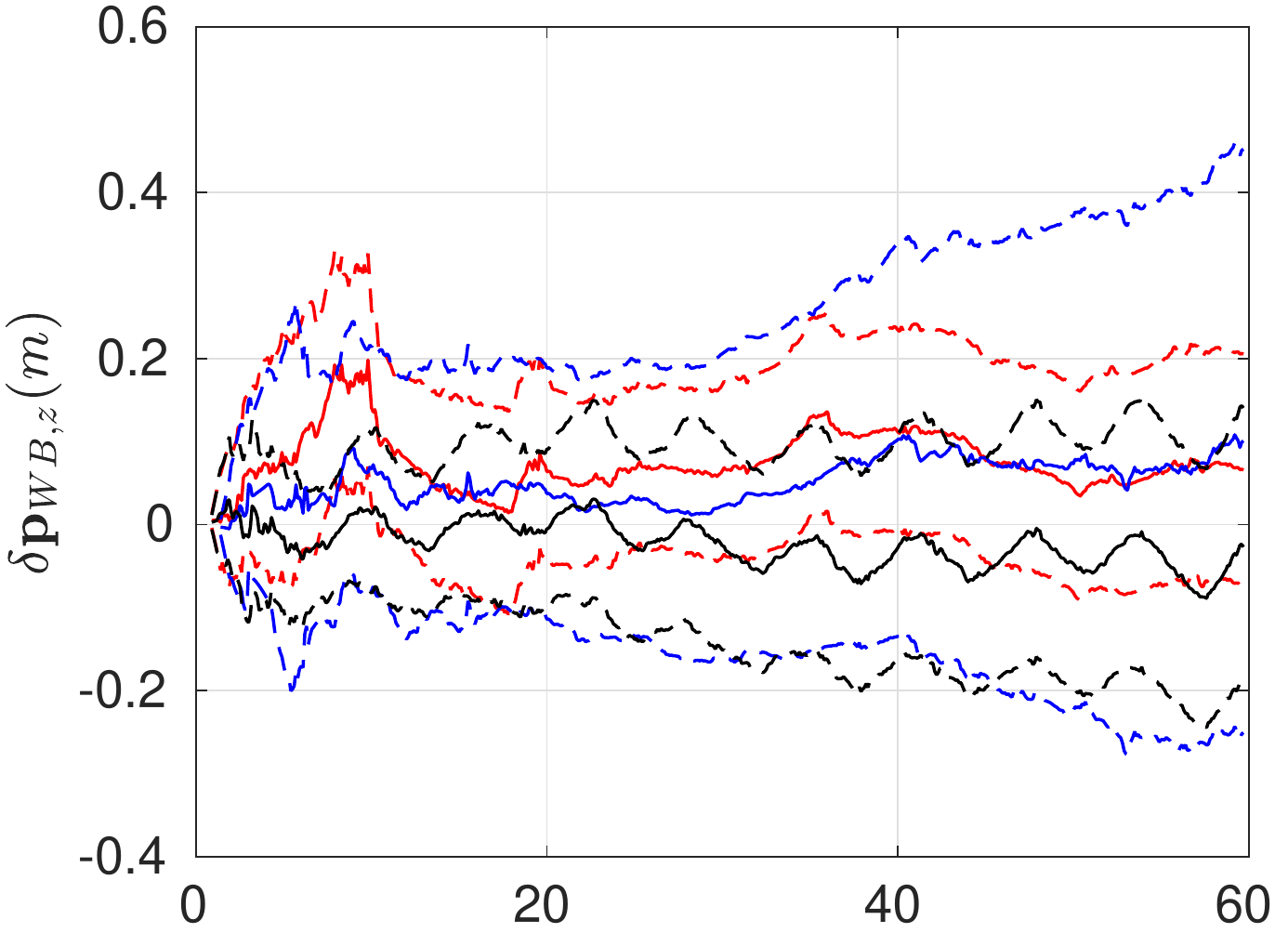}
\includegraphics[width=.32\textwidth]{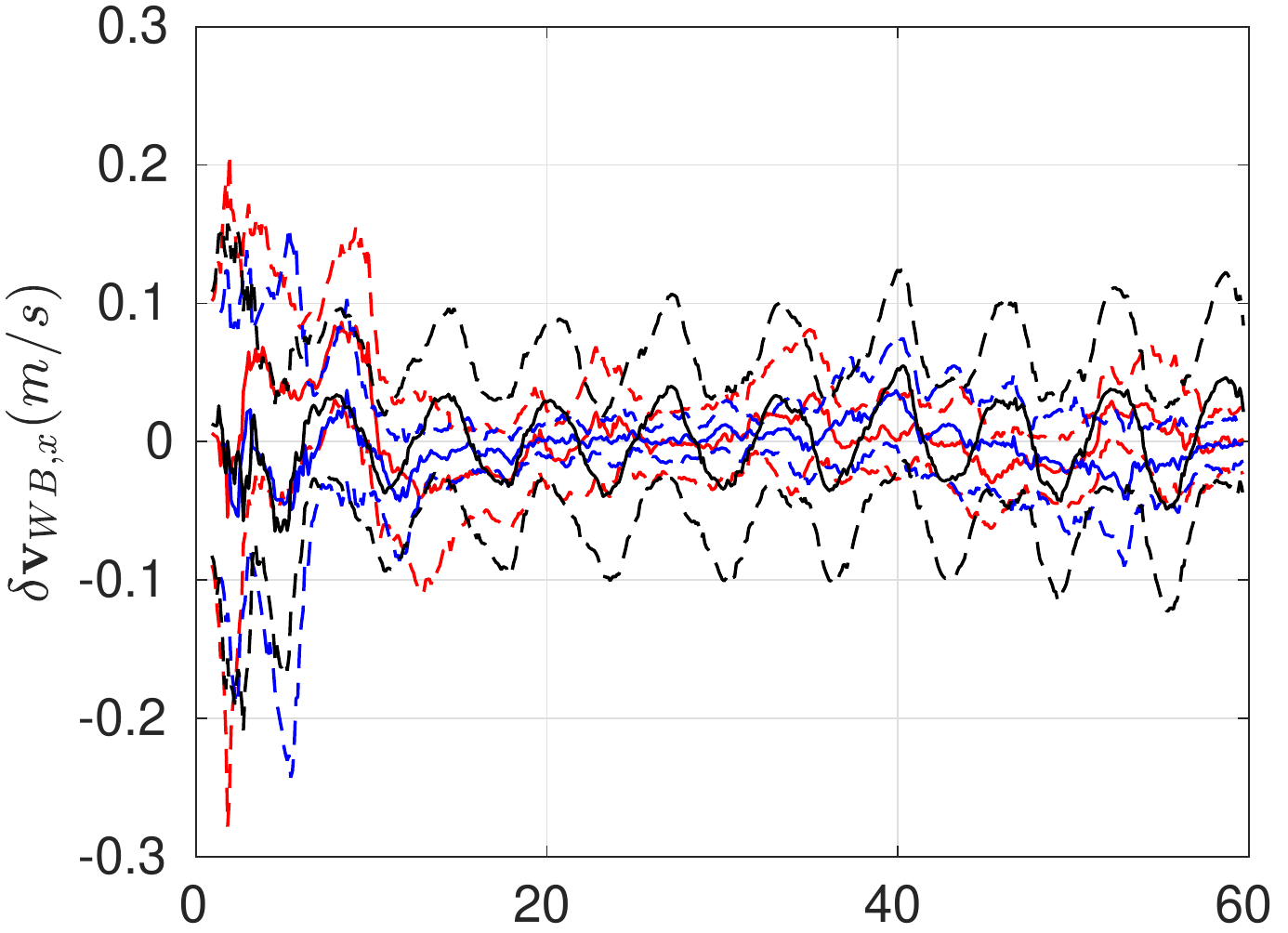}
\includegraphics[width=.32\textwidth]{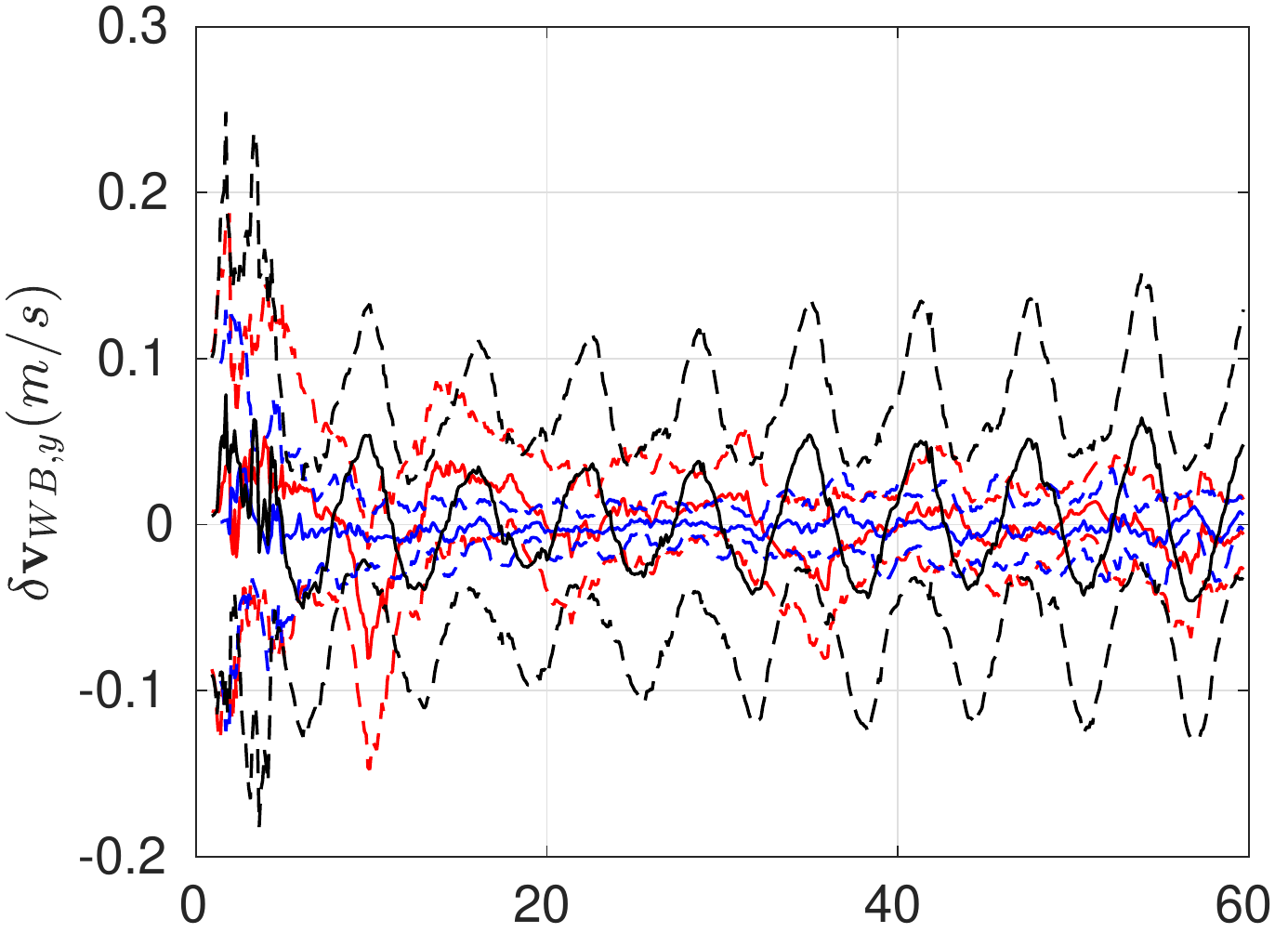}
\includegraphics[width=.32\textwidth]{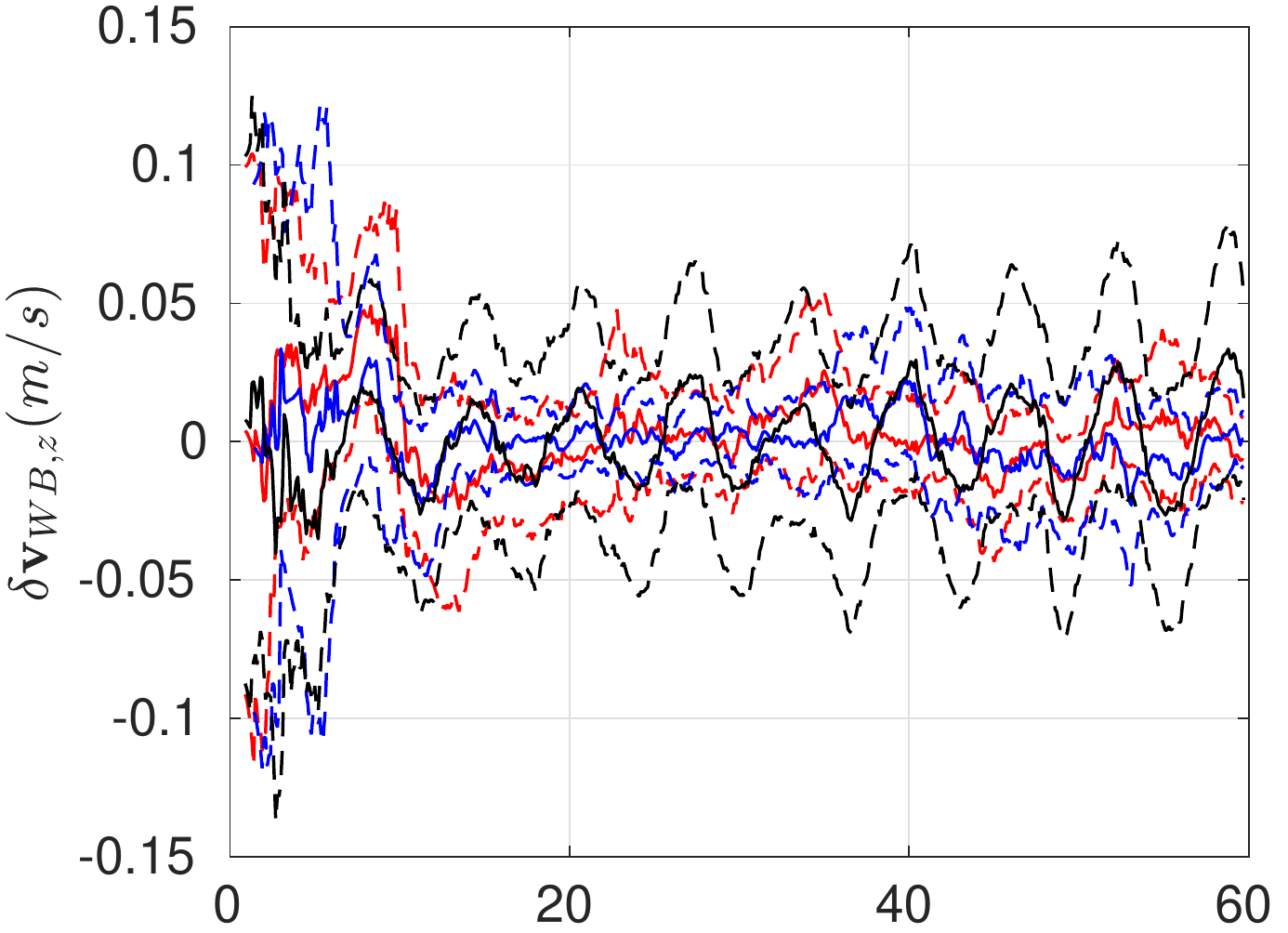}
\centering
\caption{Errors of the estimated orientation $\mbf q_{WB}$ (top row), position $\mbf p_{WB}$ (middle row), and velocity $\mbf v^{W}$ (bottom row), and their $\pm3\sigma$ bounds, under three motion types, lemniscate, line segment and circle.}
\label{fig:sv-history1}
\end{figure}

\begin{figure}[htb]
\includegraphics[width=.30\textwidth]{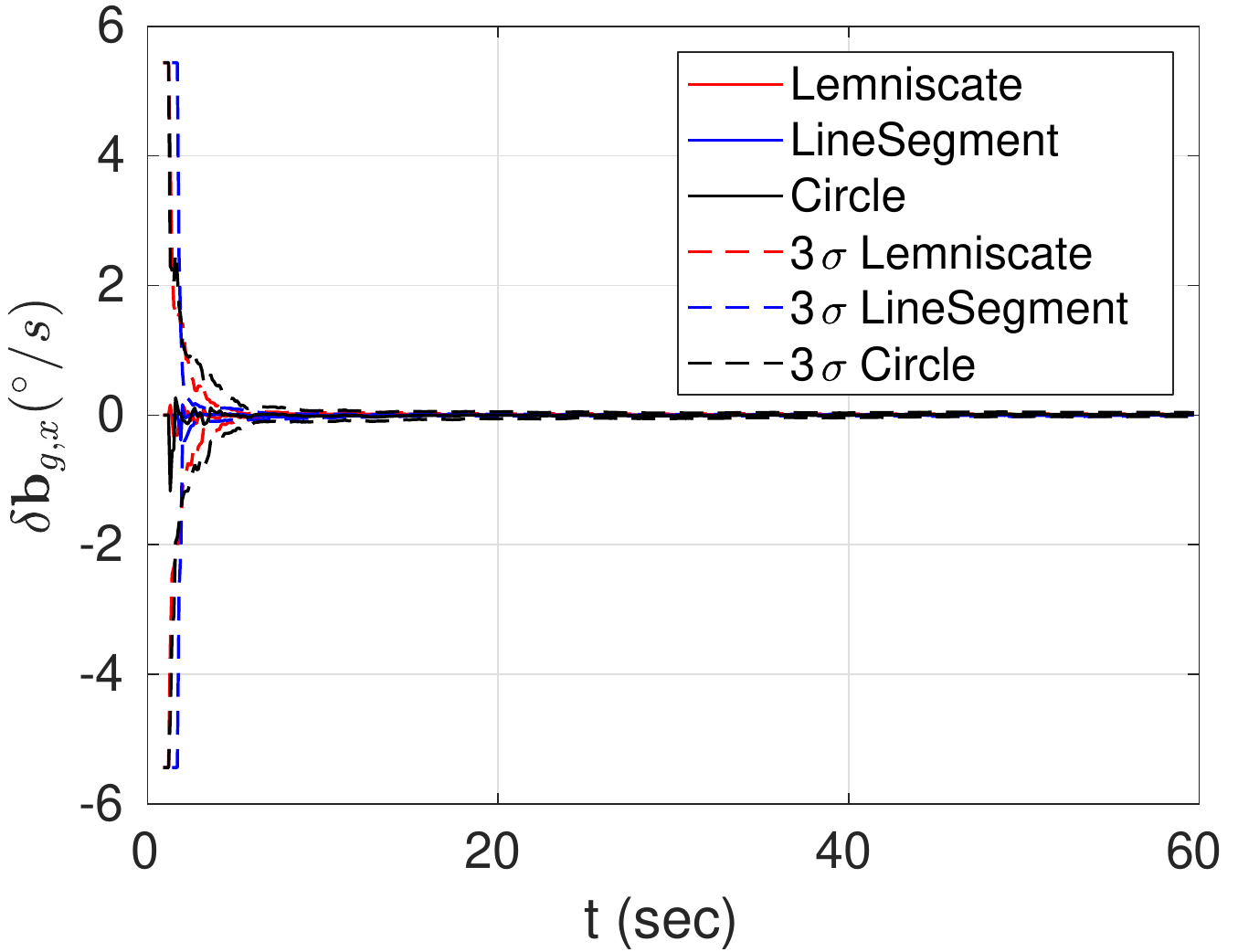}
\includegraphics[width=.30\textwidth]{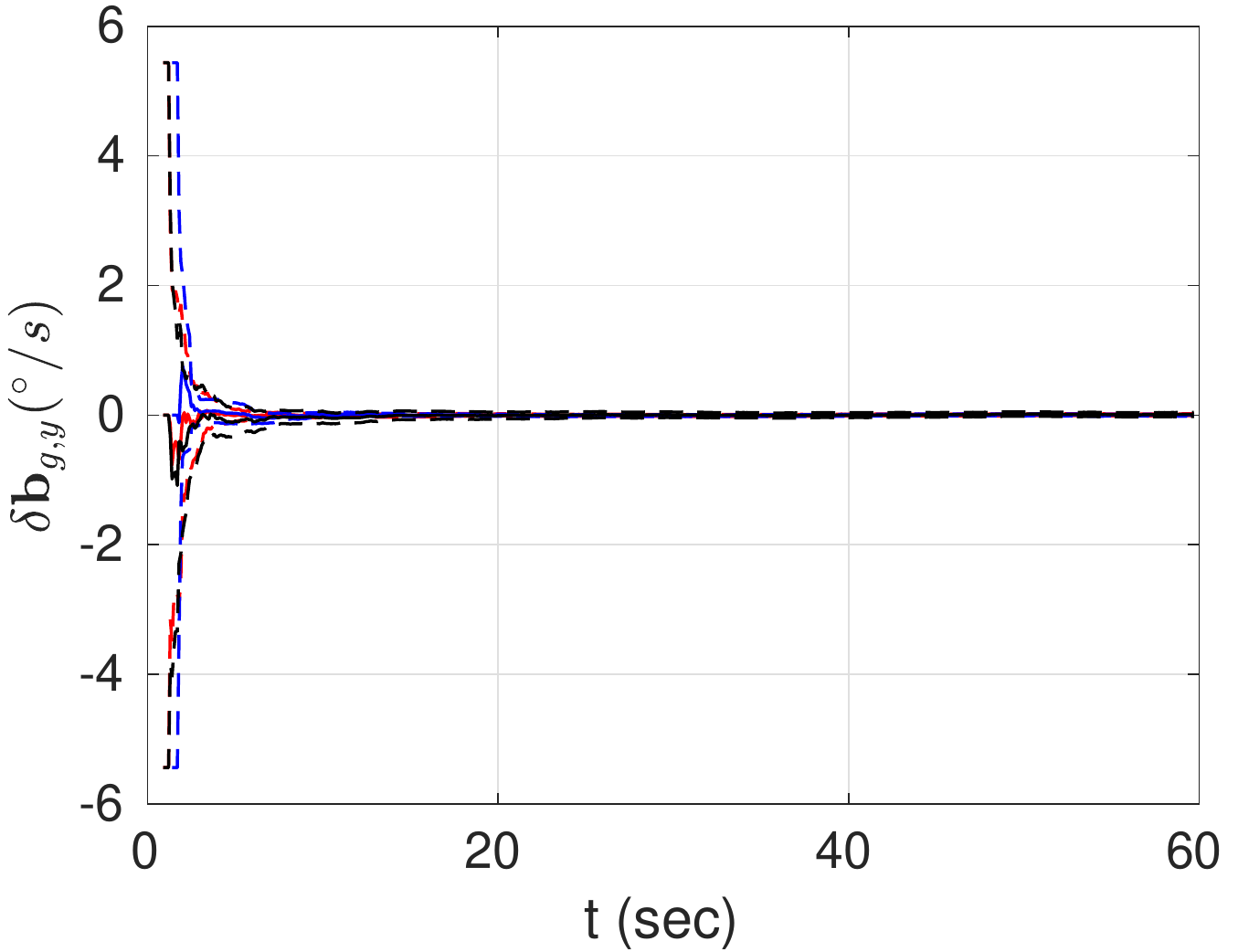}
\includegraphics[width=.30\textwidth]{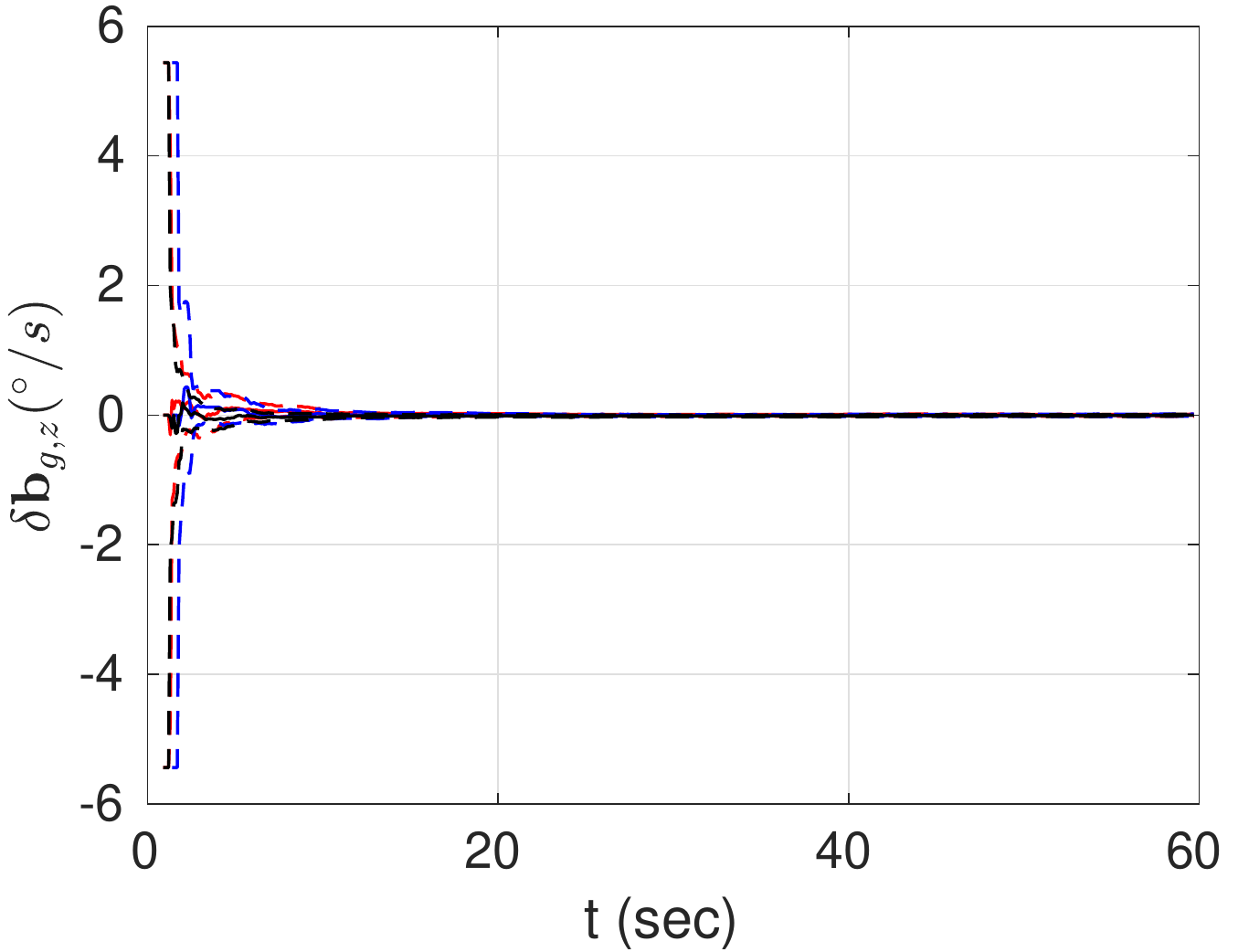}
\includegraphics[width=.32\textwidth]{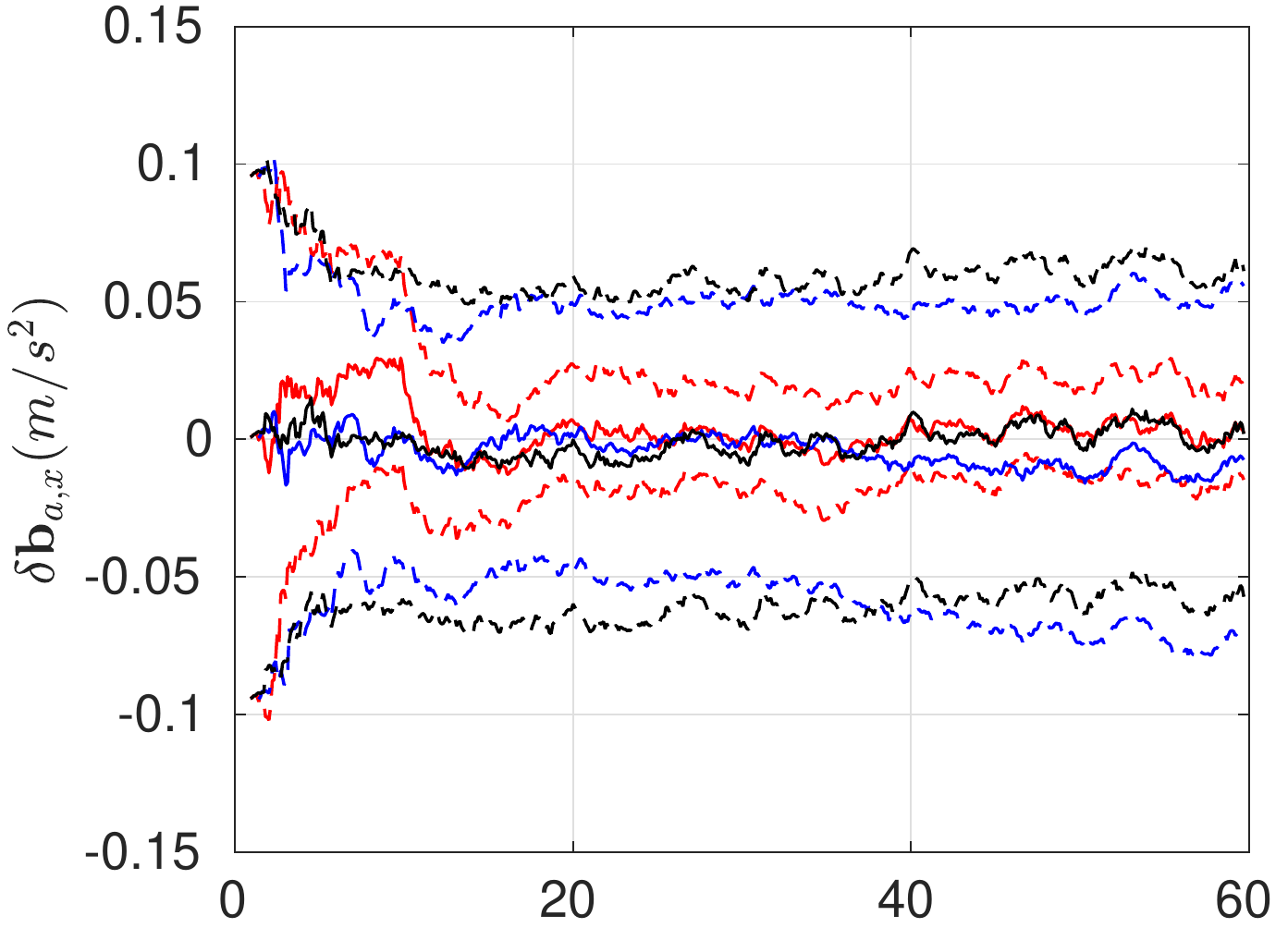}
\includegraphics[width=.32\textwidth]{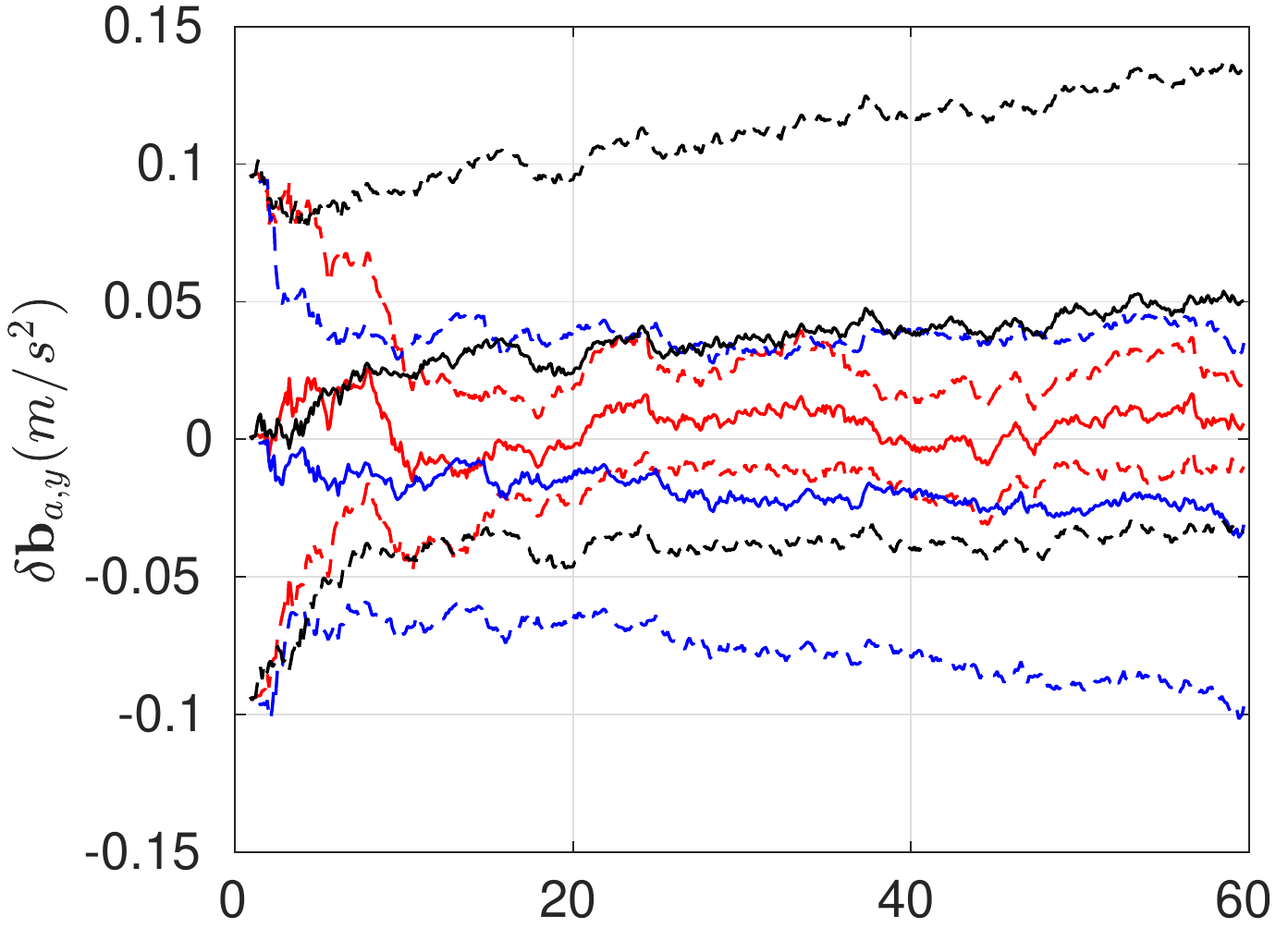}
\includegraphics[width=.32\textwidth]{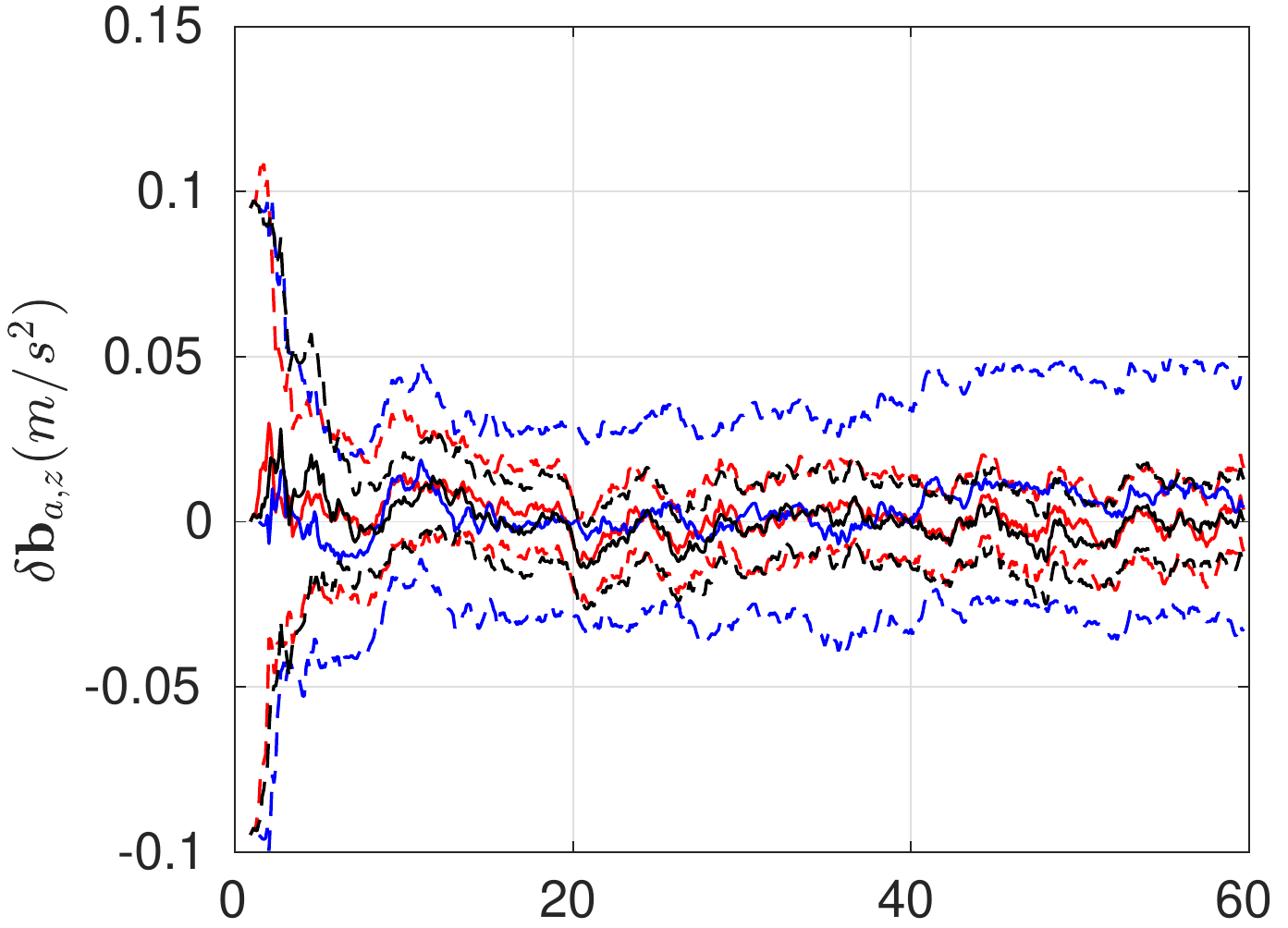}
\includegraphics[width=.30\textwidth]{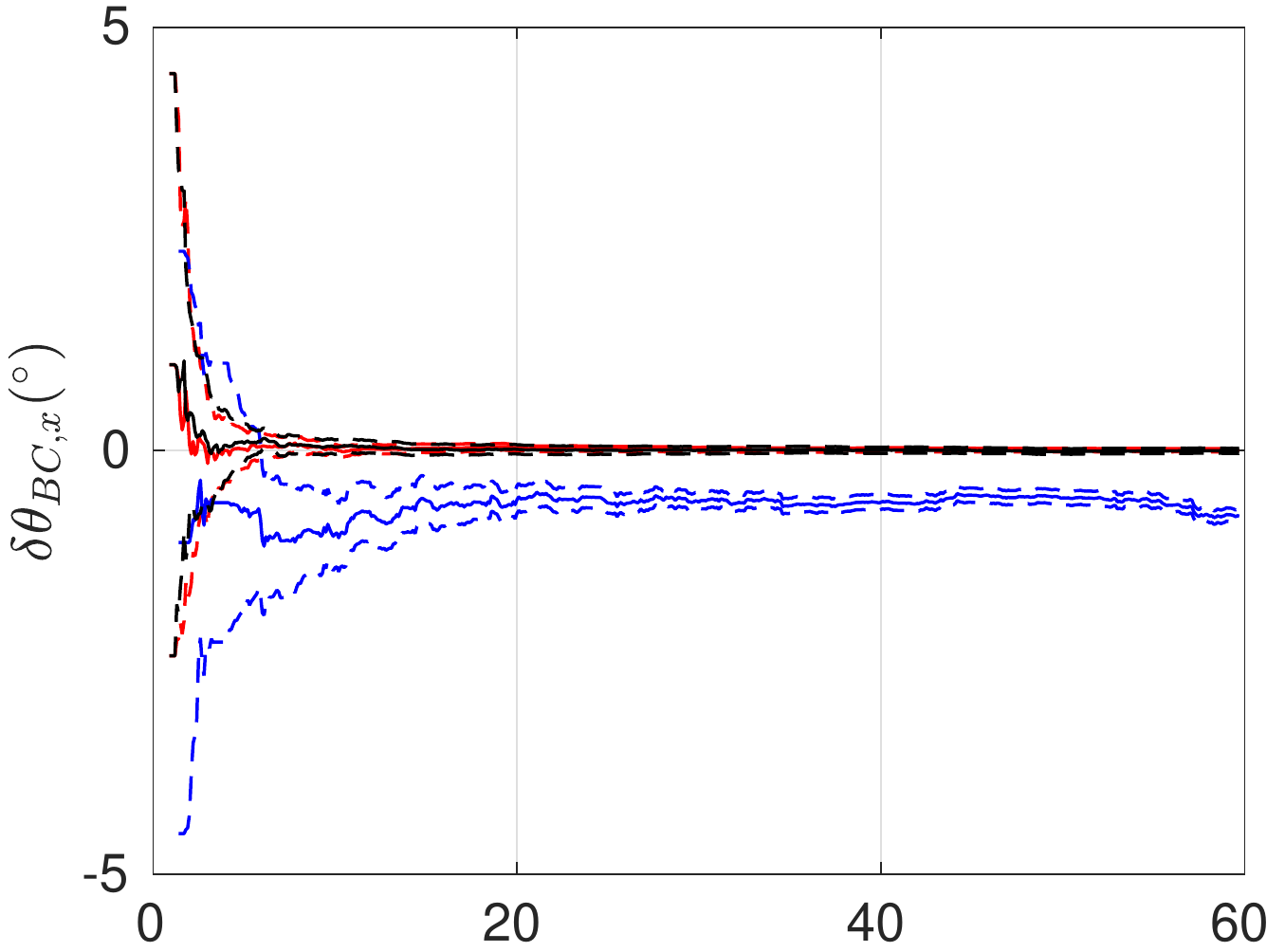}
\includegraphics[width=.30\textwidth]{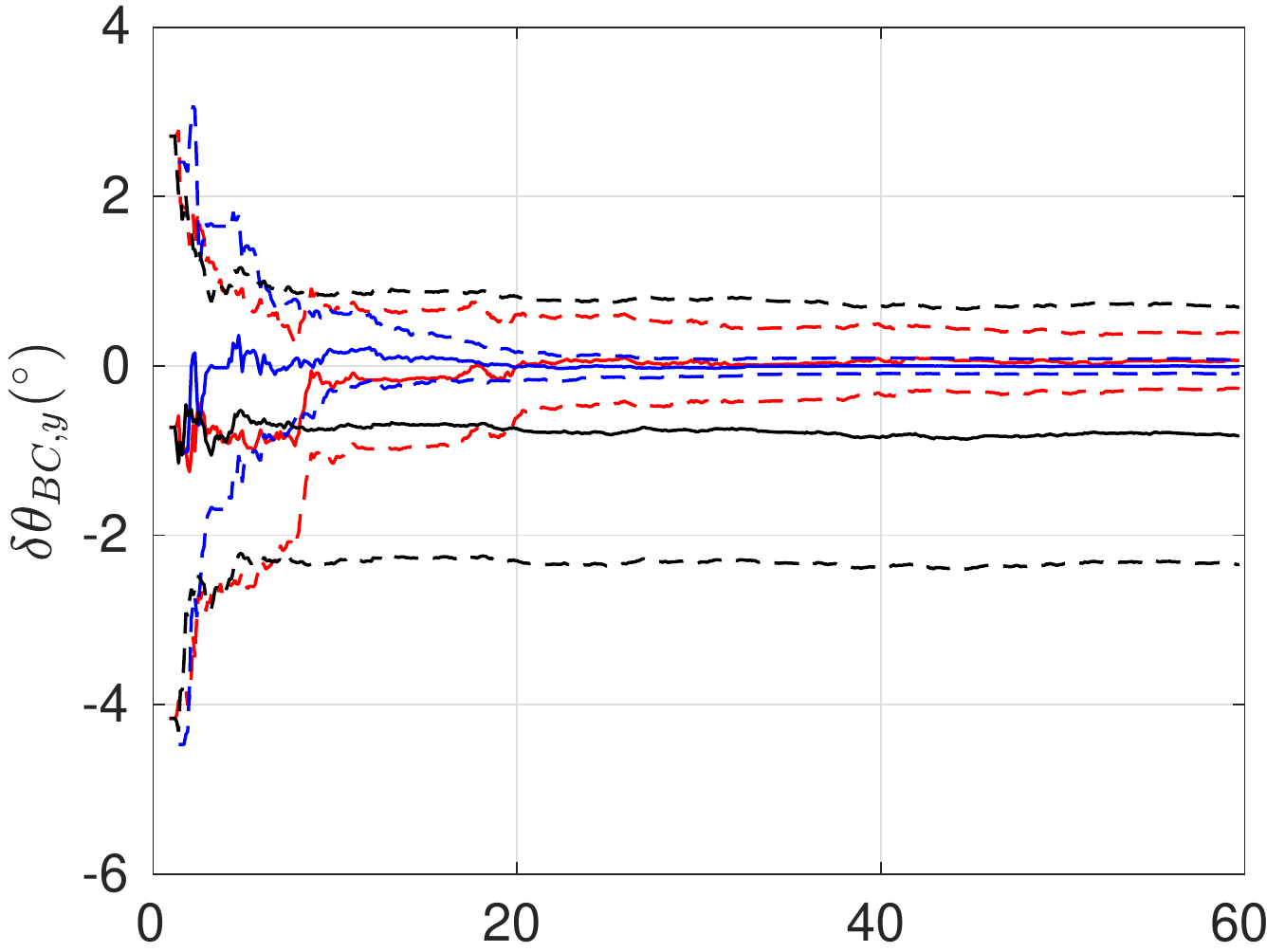}
\includegraphics[width=.30\textwidth]{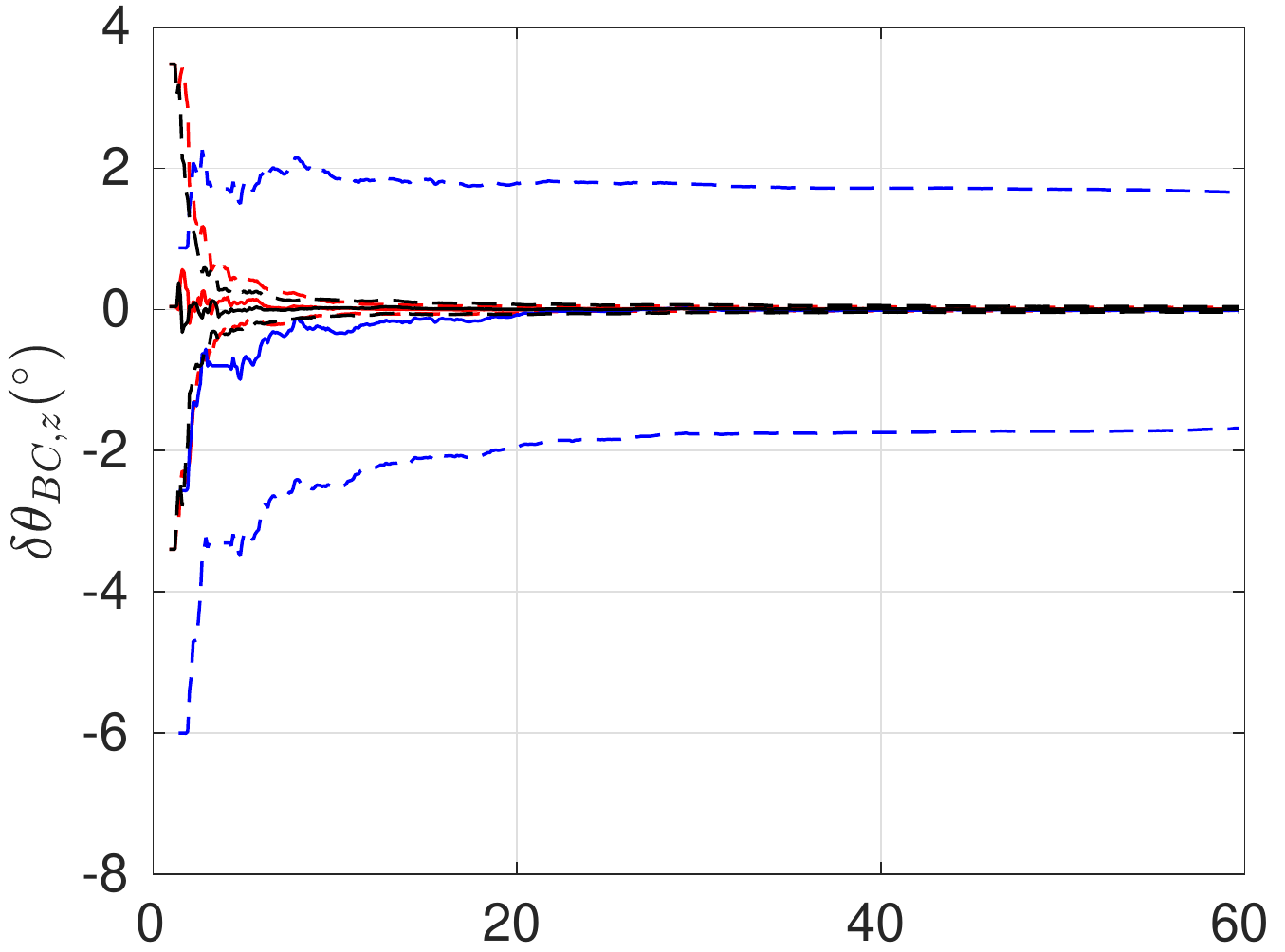}
\includegraphics[width=.24\textwidth]{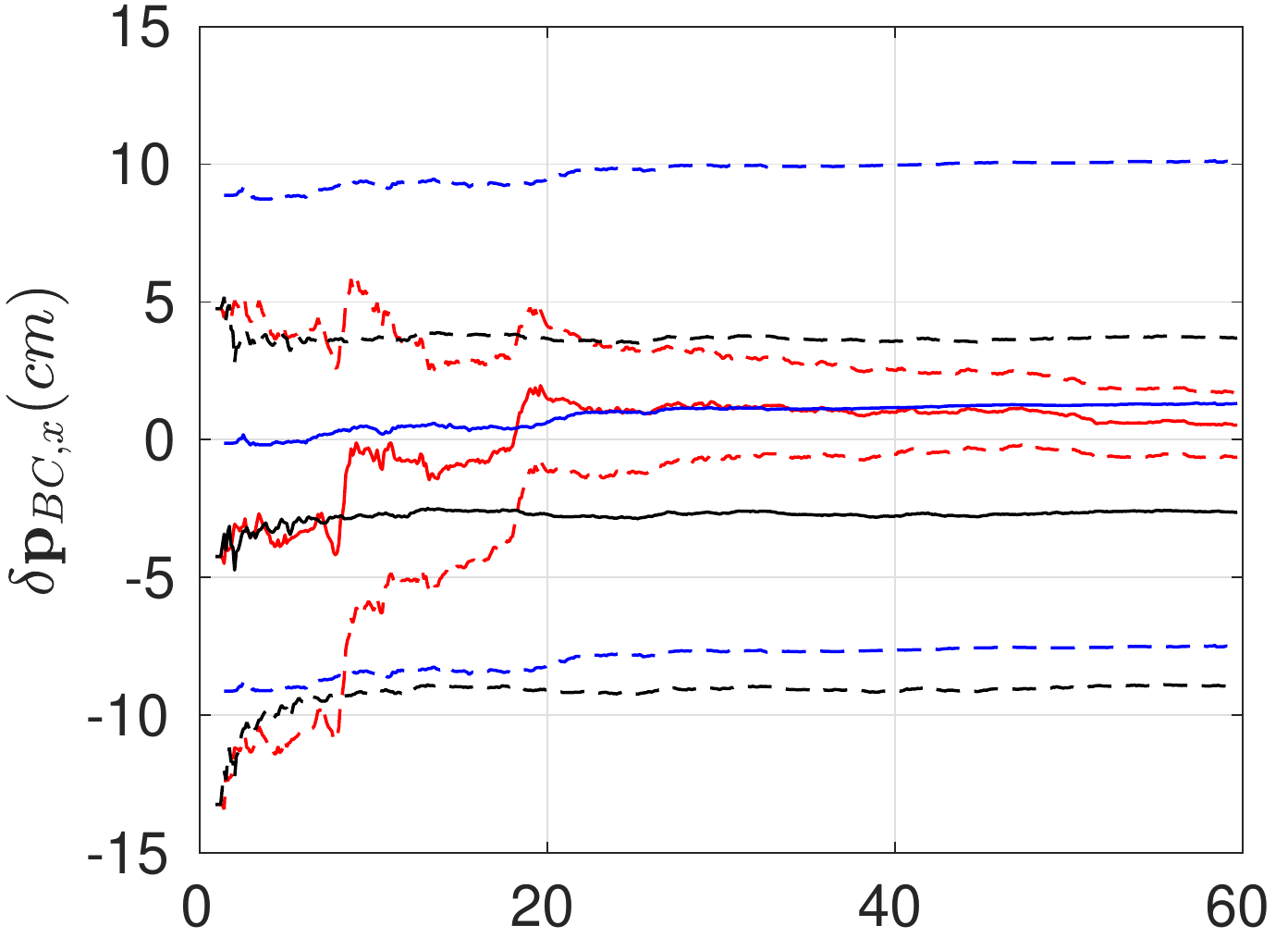}
\includegraphics[width=.24\textwidth]{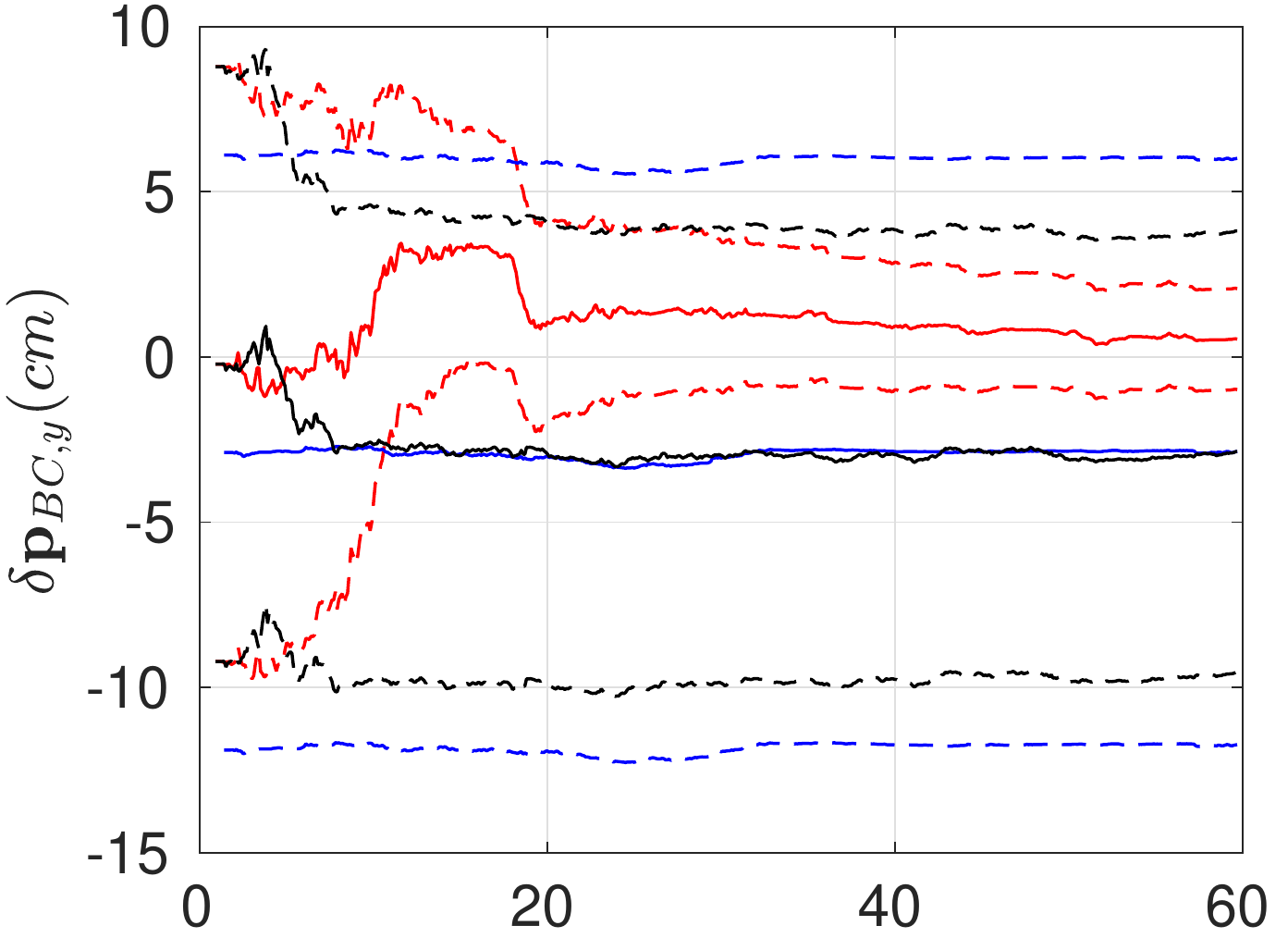}
\includegraphics[width=.24\textwidth]{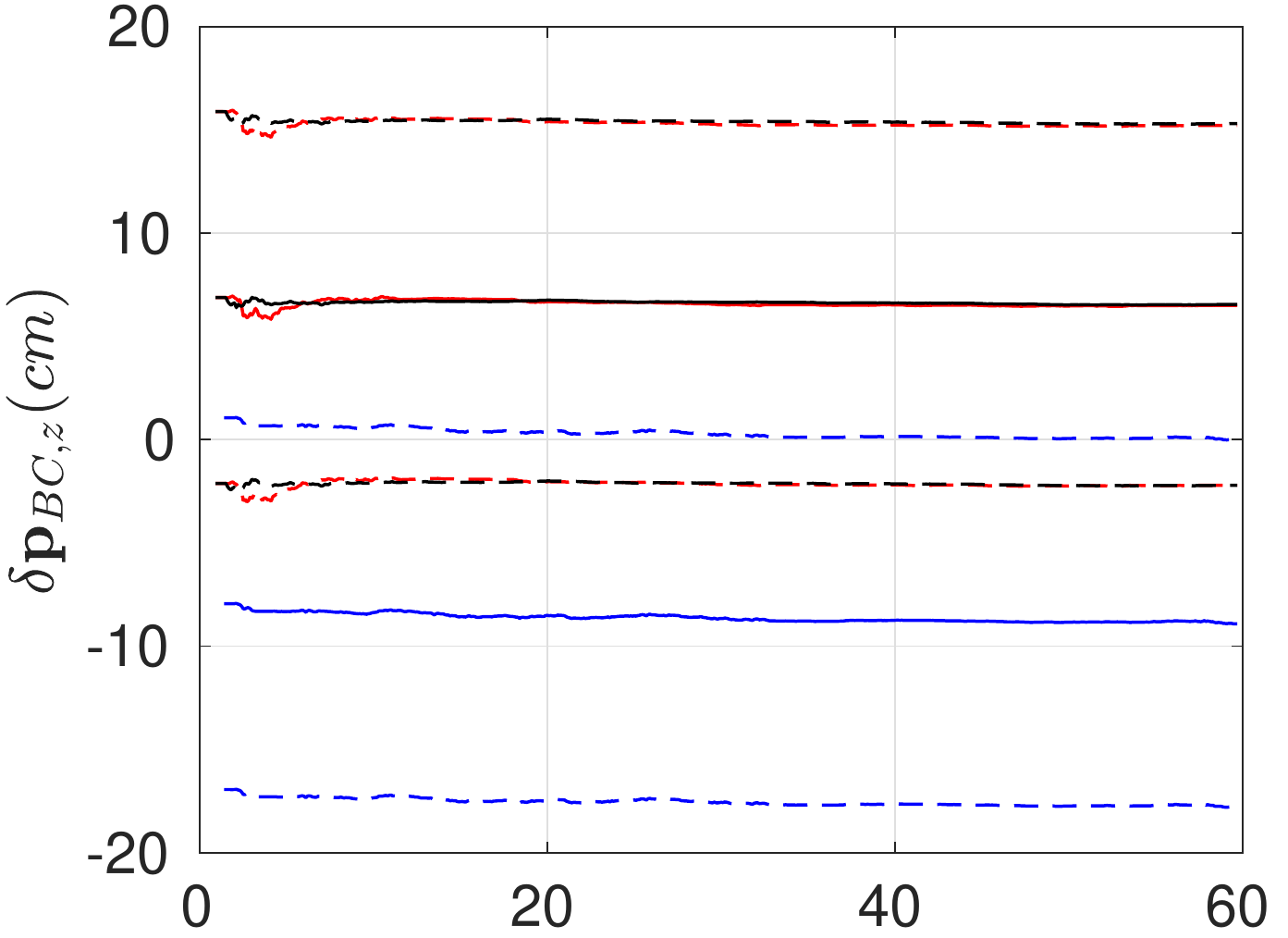}
\includegraphics[width=.24\textwidth]{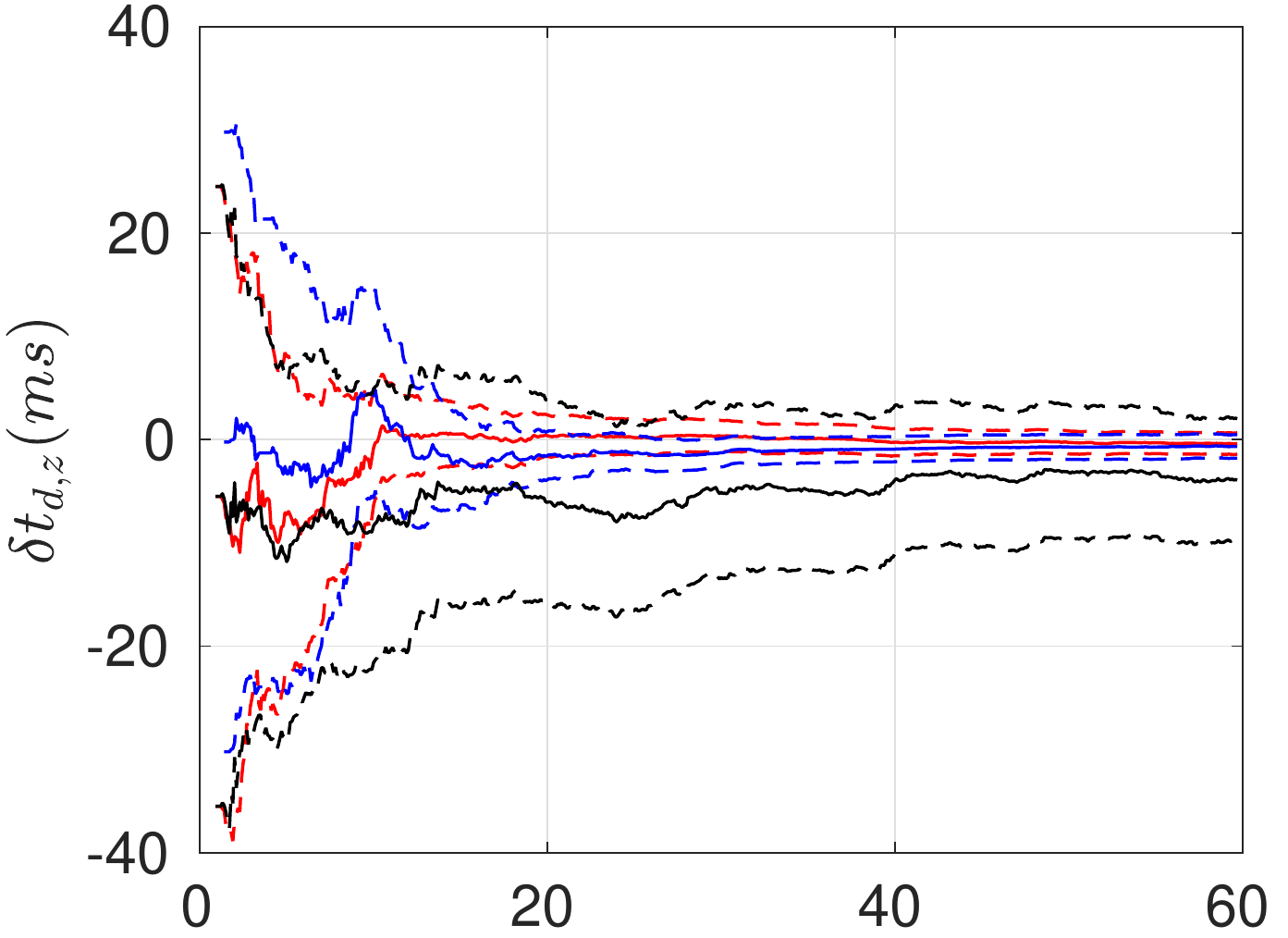}
\centering
\caption{Errors of the estimated biases $\mbf b_g$ (top row) and $\mbf b_a$ (second row), extrinsic parameters $\mbf q_{BC}$ (third row) and $\mbf p_{BC}$ (fourth row), and time offset $t_d$ (fourth row last), and their $\pm3\sigma$ bounds, under three motion types, lemniscate, line segment and circle.}
\label{fig:sv-history2}
\end{figure}

The estimated state variables are 
\begin{equation}
\label{eq:deg-motion-vars}
\{\mbf p_{WB}, \mbf{q}_{WB}, \mbf{v}^W, \mbf{b}, \mbf p_{BC}, \mbf{q}_{BC}, t_d\}.    
\end{equation}
Note that OpenVINS estimates a slightly different set of variables, we convert their estimates and standard deviations to those of \eqref{eq:deg-motion-vars} for result presentation.
The estimation errors and standard deviations for \eqref{eq:deg-motion-vars} in a typical run under these planar trajectories are depicted in Figs.~\ref{fig:sv-history1} and \ref{fig:sv-history2}.

From Fig.~\ref{fig:sv-history1}, we see that: (1) With lemniscate and circle trajectories, the yaw component of $\mbf q_{WB}$ does not converge. For the line segment motion, all three components of $\mbf q_{WB}$ show slightly widening $3\sigma$ bounds, implying that
$\mbf q_{WB}$ is unobservable with only translation.
(2) For all three trajectories, components of $\mbf p_{WB}$ have growing errors and widening $3\sigma$ bounds, and components of $\mbf v^W$ have bounded errors and $3\sigma$ curves.

From Fig.~\ref{fig:sv-history2}, it can be seen that: (1) Gyroscope biases converge well with all three motion types. (2) For the line segment trajectory, accelerometer biases do not converge, and for the circle motion, the x- and y- components of the accelerometer biases do not converge, as $\mbf b_a$ is indeterminable in both cases.
(3) For the line segment motion, the z-component of $\mbf q_{BC}$ has standard deviations that only slightly tighten over time, similarly to the y-component of $\mbf q_{BC}$ with the circle motion. Moreover, the x-component of $\mbf q_{BC}$ converges to a value of some offset with line segment motion. These effects are not predicted by the observability analysis and are likely due to that the filter estimating unobservable parameters tends to be trapped in local minima.
(4) For both line segment and circle motion, all three components of $\mbf p_{BC}$ have non-decreasing errors and standard deviations. For the lemniscate motion, $\mbf p_{BC}$ converges except for the component along the rotation axis (z-axis of the IMU frame), concurring with the analysis results.
(5) In contrast to lemniscate and line segment motion, the time offset with the circle motion converges to a slightly wrong value and have relatively large variances, pointing to the unobservable property of time offset with motion of constant local acceleration.

\subsection{Unobservable Time Offset with Constant Control Inputs}
\label{subsec:exp-time-offset}
\begin{figure}[htb]
\includegraphics[width=.45\textwidth]{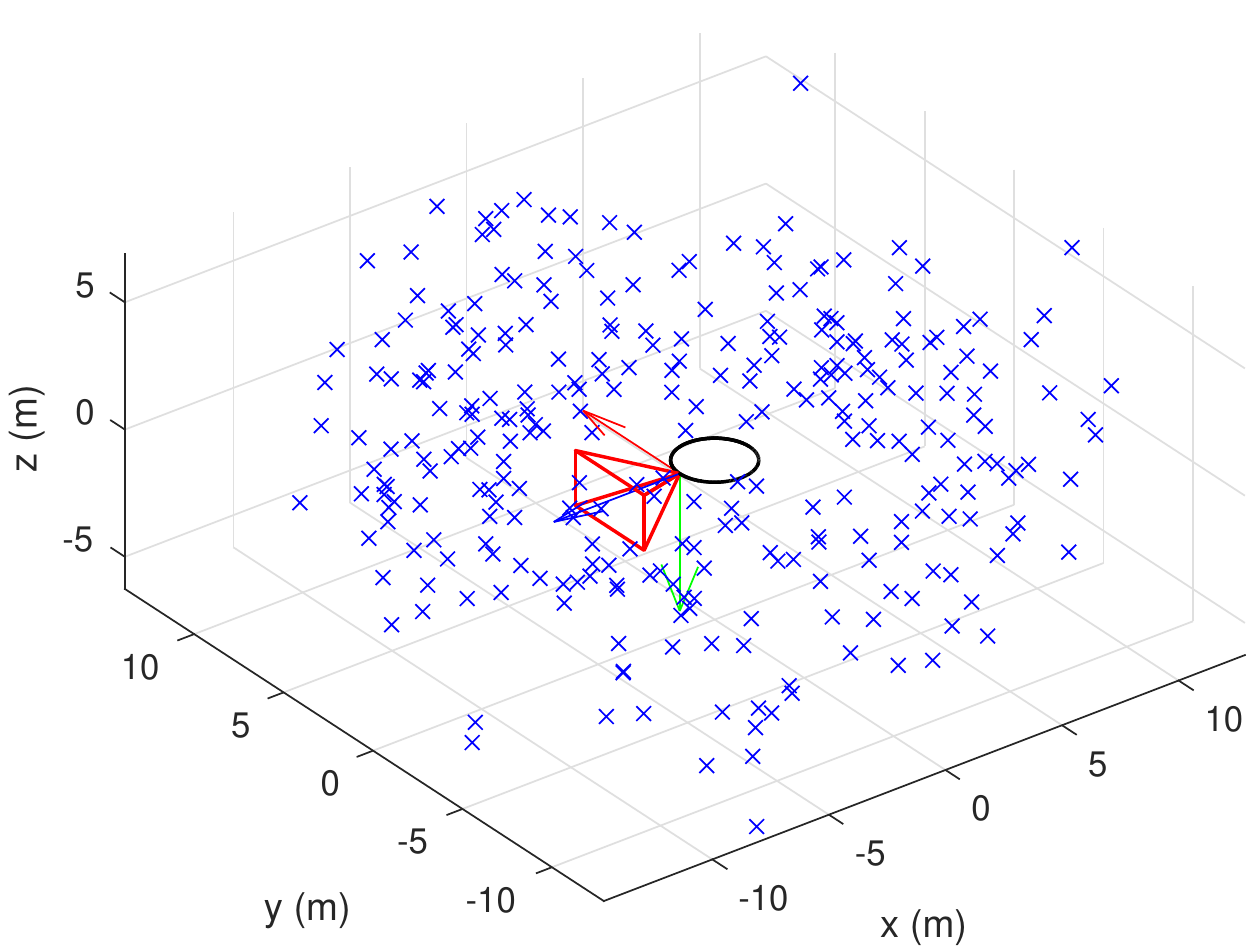}
\includegraphics[width=.4\textwidth]{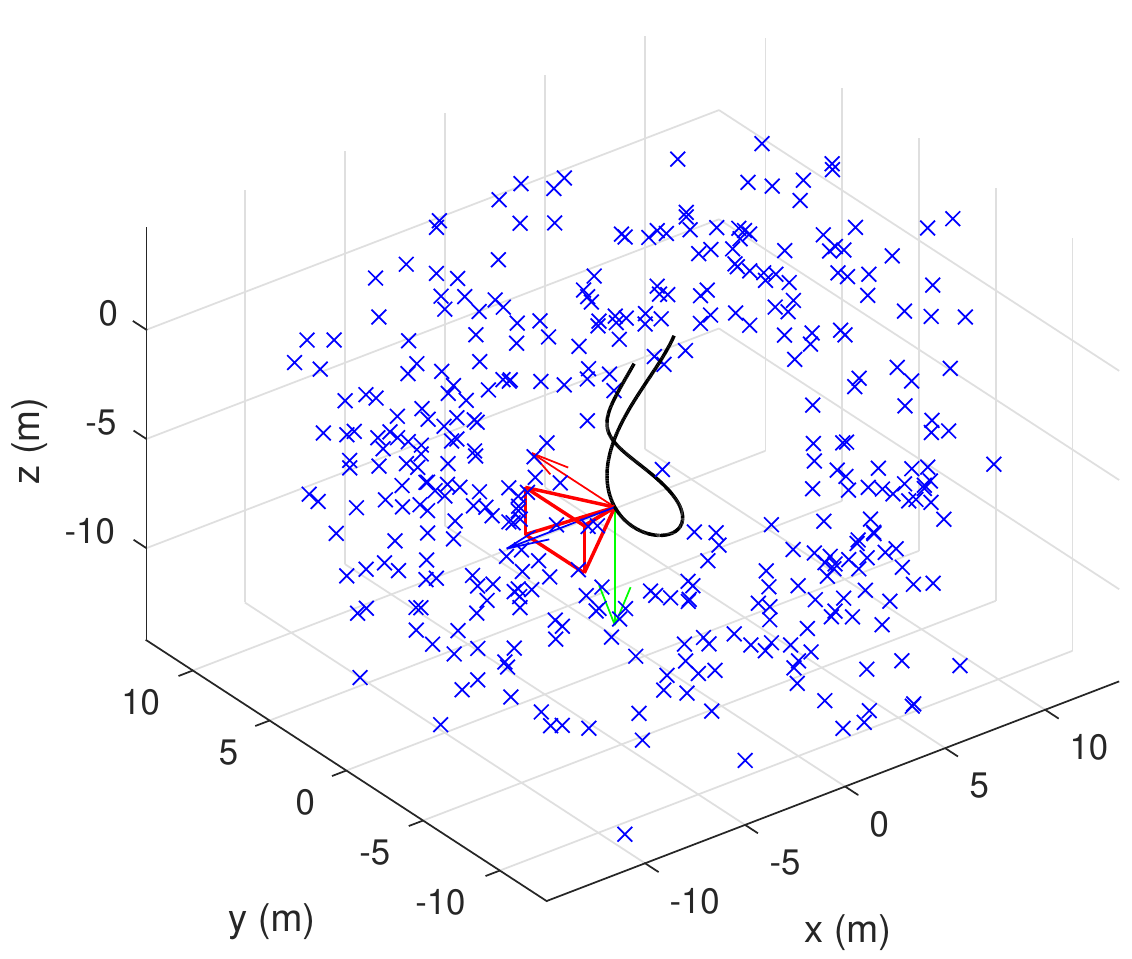}
\includegraphics[width=.45\textwidth]{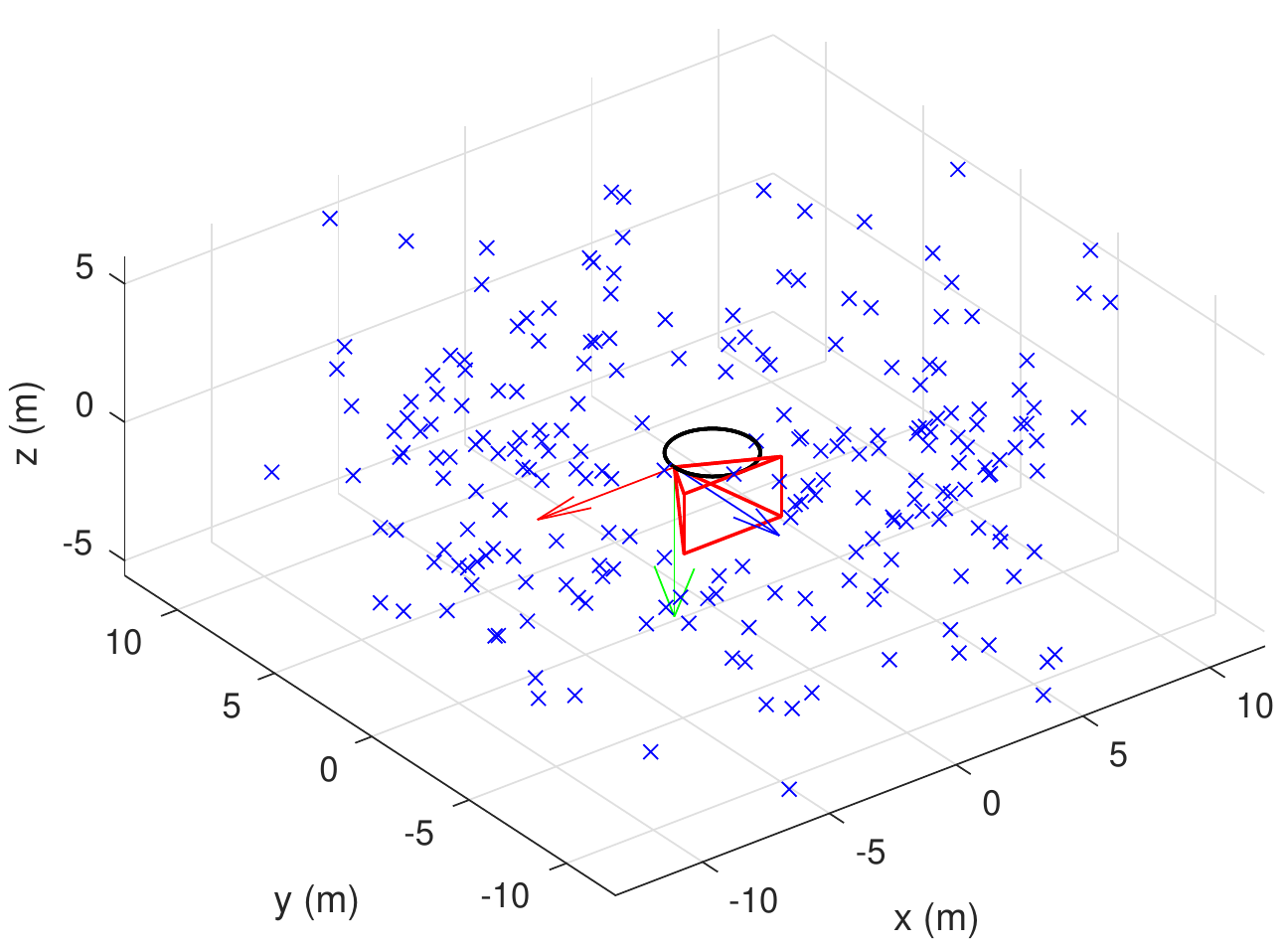}
\includegraphics[width=.45\textwidth]{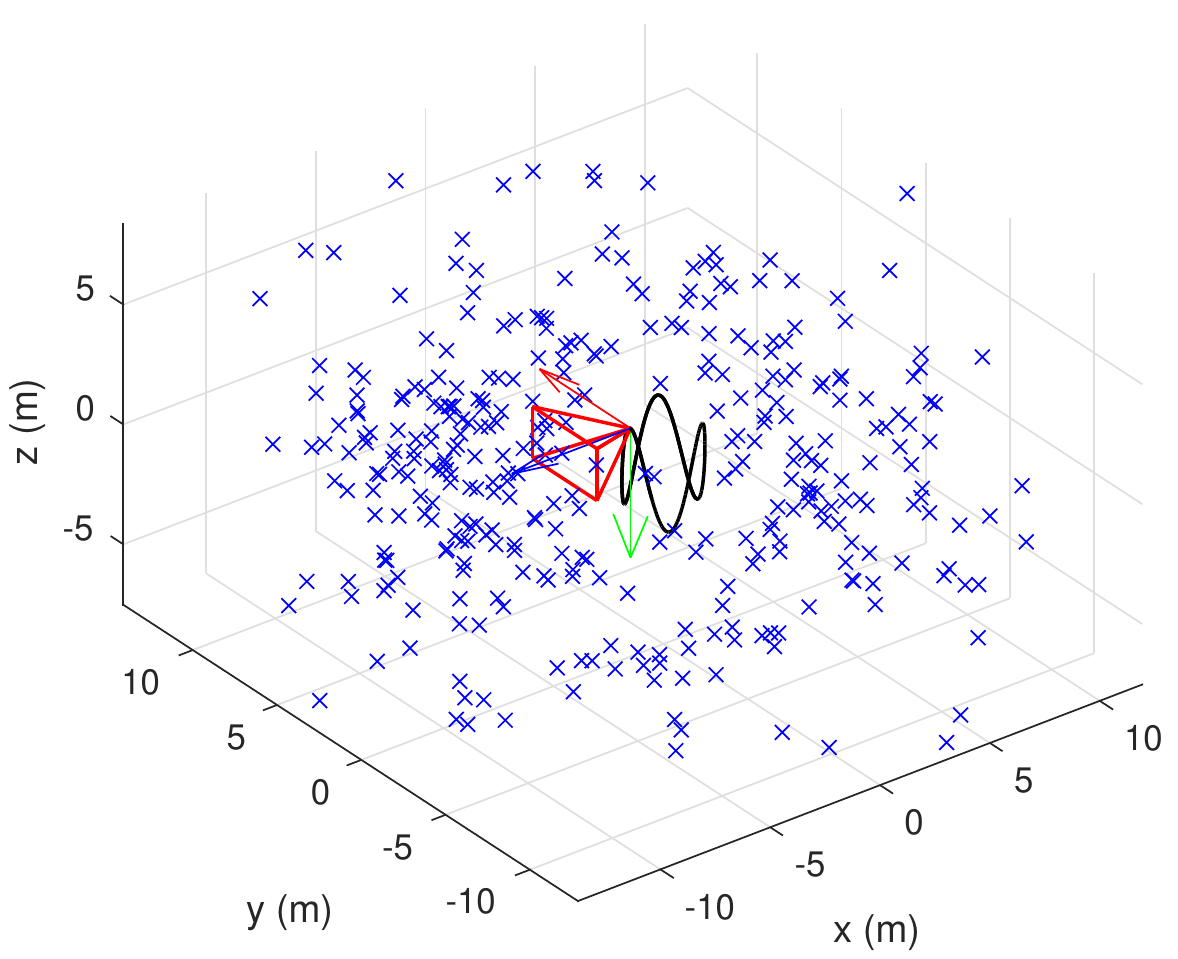}
\centering
\caption{Degenerate motion types and the simulated landmarks: (top left) circular motion with constant local angular and linear velocity, (top right) cylindrical motion with constant local angular velocity and constant local linear acceleration, (bottom left) circular motion with varying local angular velocity and varying local linear acceleration, (bottom right) cylindrical motion with constant local angular velocity and varying local (vertical) linear acceleration. They correspond to cases a, b, c, and d in Section~\ref{subsec:exp-time-offset}.}
\label{fig:degenerate-motion}
\end{figure}

To validate the observability property of the time offset in a VIO system,
we choose four motion types: (a) circular motion with constant local angular velocity $\mbs \omega_{WB}^B$ and constant local linear velocity $\mbf v^B$, (b) cylindrical motion with constant local angular velocity $\mbs \omega_{WB}^B$ and constant local linear acceleration $\mbf a^B$, (c) circular motion with varying local angular velocity $\mbs \omega_{WB}^B$ and hence varying local linear acceleration $\mbf a^B$, (d) cylindrical motion with constant local angular velocity $\mbs \omega_{WB}^B$ and varying local linear acceleration $\mbf a^B$, shown in Fig.~\ref{fig:degenerate-motion}.
Based on our analysis, it is expected that the time offset is unobservable for cases a and b, and is observable for cases c and d.
In contrast, according to the linearized model (Table~\ref{tab:time-offset-obs-cond}), the time offset is unobservable for case a, and is observable for cases b, c, and d.

\begin{figure}[htb]
\includegraphics[width=.45\textwidth]{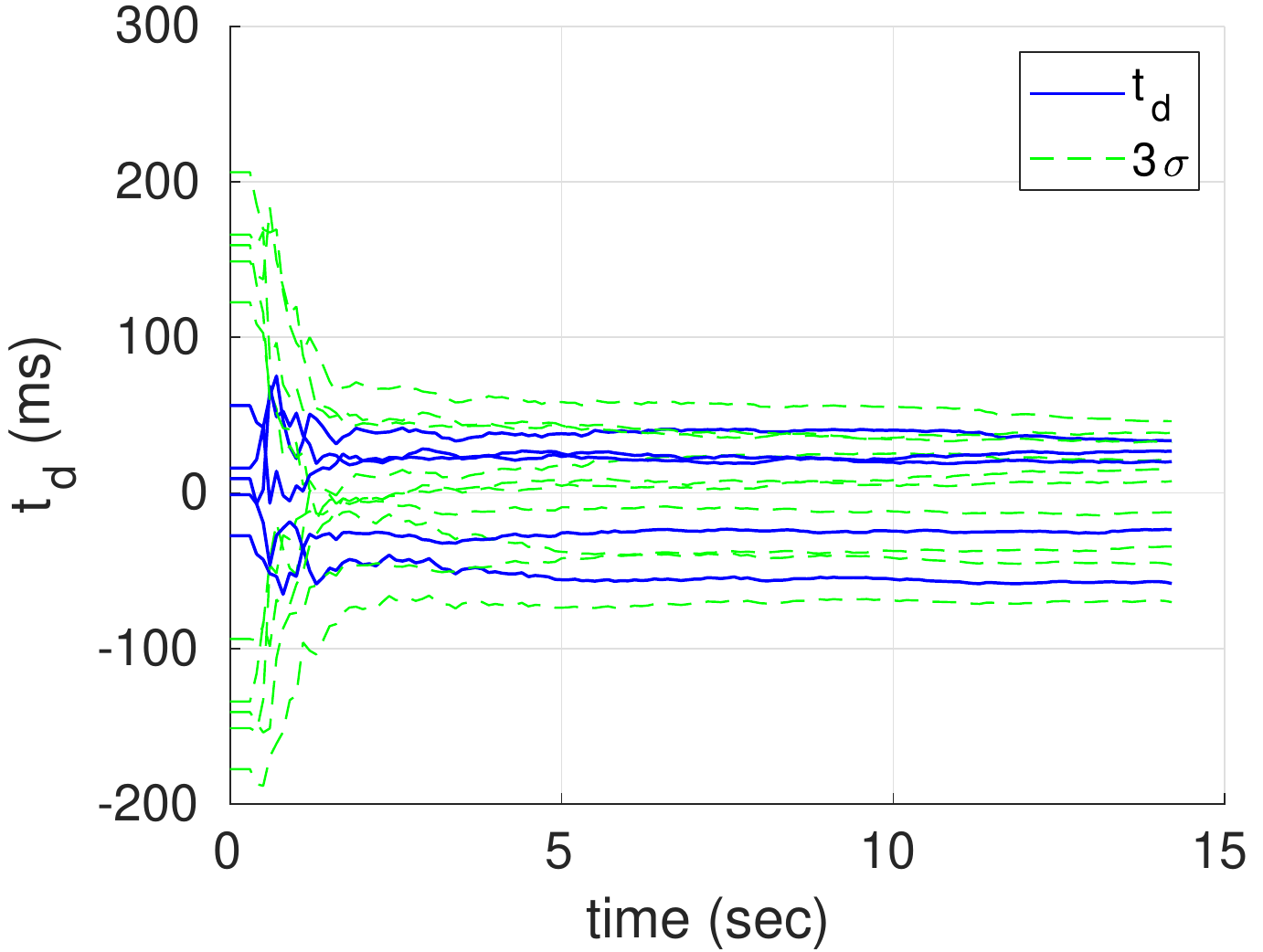}
\includegraphics[width=.45\textwidth]{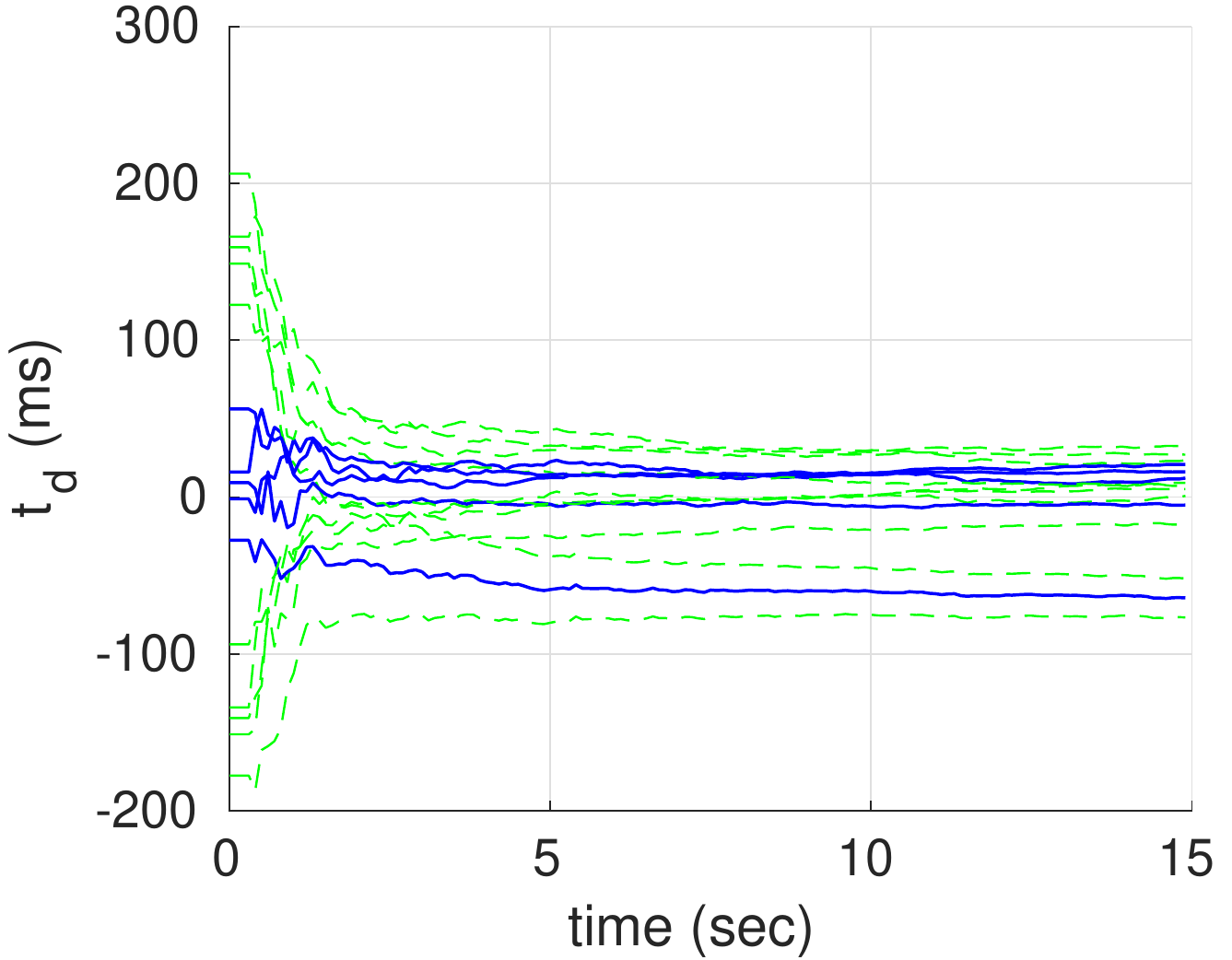}
\includegraphics[width=.45\textwidth]{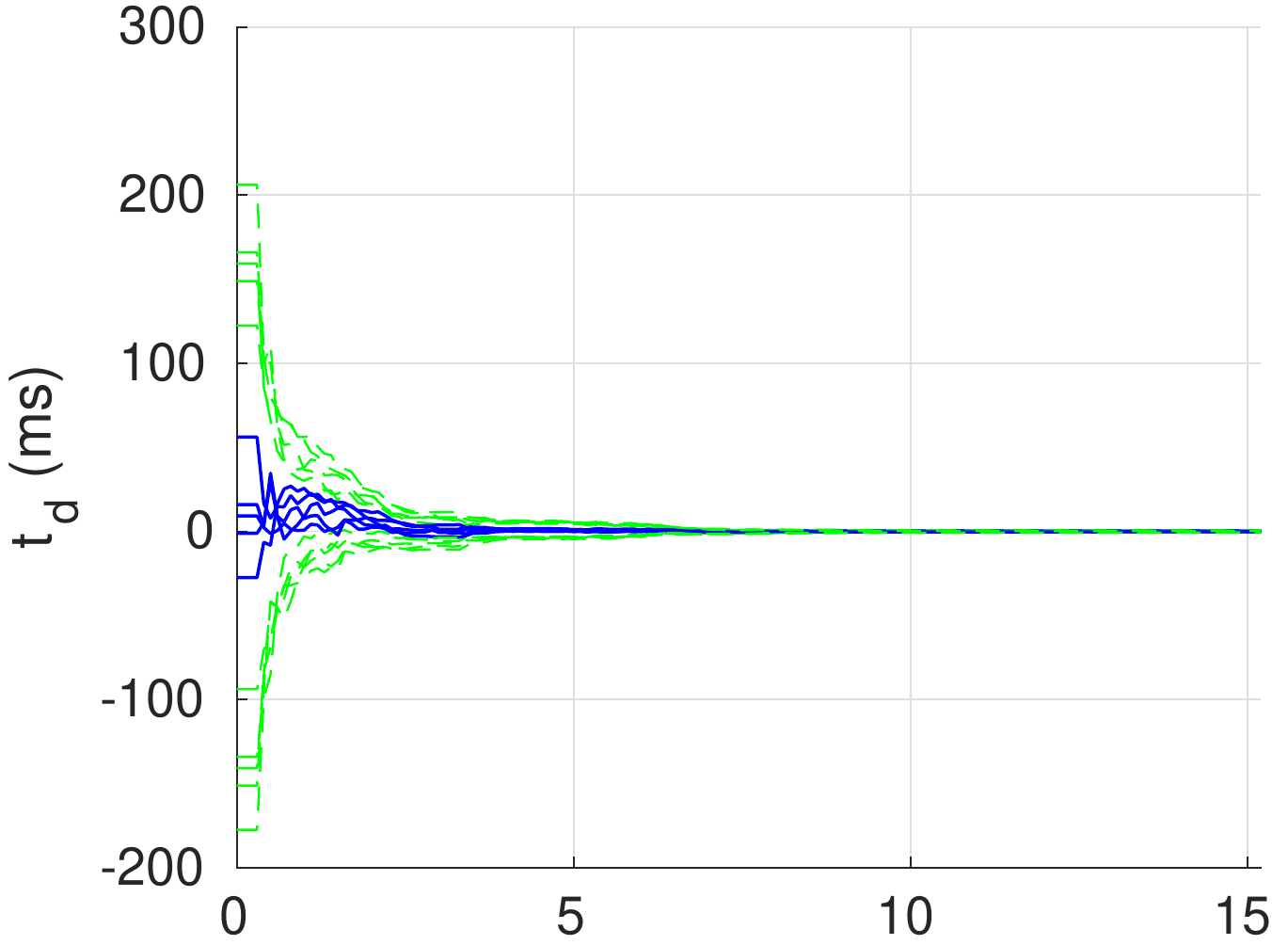}
\includegraphics[width=.45\textwidth]{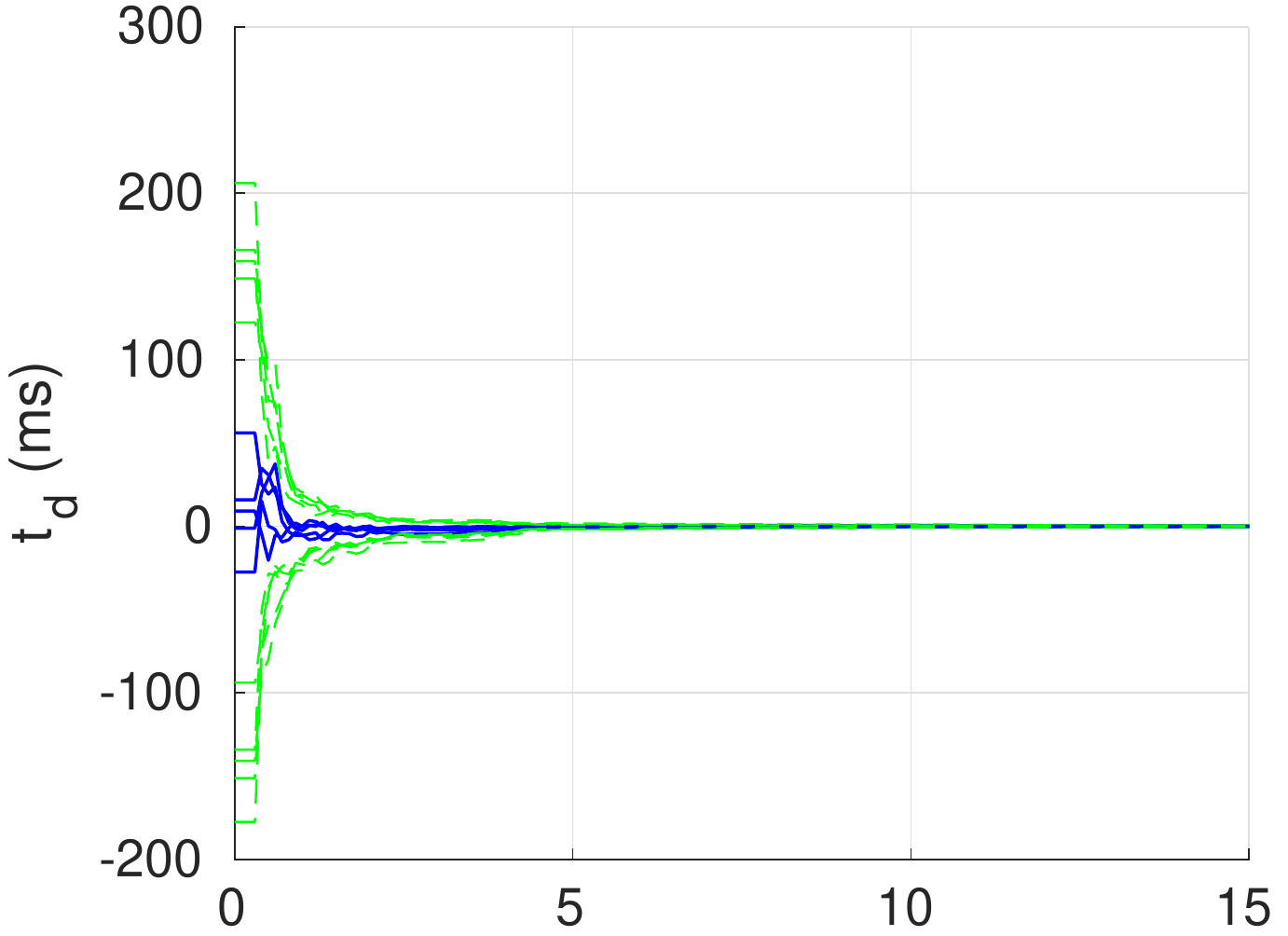}
\centering
\caption{Estimated time offset and $\pm 3\sigma$ envelopes of five runs on each motion type with different initial time offsets. The true time offset is 0. The motion types are: (top left) circular motion with constant local angular and linear velocity, (top right) cylindrical motion with constant local angular velocity and constant local linear acceleration, (bottom left) circular motion with varying local angular velocity and varying local linear acceleration, (bottom right) cylindrical motion with constant local angular velocity and varying local (vertical) linear acceleration. They correspond to cases a, b, c, and d in Section~\ref{subsec:exp-time-offset}.}
\label{fig:timeoffset}
\end{figure}

The estimated state variables by the OpenVINS filter are 
\begin{equation}
\label{eq:time-offset-vars}
\{\mbf p_{WB}, \mbf{q}_{WB}, \mbf{v}^W, \mbf{b}, t_d\},
\end{equation}
where the extrinsic parameters are excluded to reduce their side effects.
Before each run of the VIO filter, a simulated time offset was sampled from a Gaussian distribution $N(0, 0.05\enskip\mathrm{sec})$, and deducted from true timestamps of the camera frames.
We repeated the simulation five times for every trajectory.
The estimated time offsets and $3\sigma$ envelopes for five runs of each degenerate motion are shown in Fig.~\ref{fig:timeoffset}.
From the top right plot of Fig.~\ref{fig:timeoffset}, we see that the time offset is unobservable with constant local angular rate and constant local linear acceleration, validating that our identified conditions when the time offset is unobservable are broader than those identified with the linearized model.
The bottom two plots of Fig.~\ref{fig:timeoffset} show that varying local angular velocity or varying local linear acceleration will render time offset observable, implying that the constant control inputs are almost necessary to make the time offset unobservable.

In summary, the simulation results confirm the analysis findings and show the effectiveness of the proposed conversion procedures.

\section{Conclusions and Future Work}
Based on the well-grounded observability analysis with Lie derivatives,
this paper investigates the observability of a nonlinear system affine in inputs when extra constraints on the control inputs or observations are applied.
We present detailed procedures to convert the nonlinear system with constraints affine in control inputs to a constraint-free standard form which is ready for automated deduction.
By using the proposed procedures, we analyze the observability properties of a VIO system under constraints on control inputs.
The analysis results about observability of VIO with constraints and time offset coincide or extend existing reports, and are validated by simulation.
Moreover, the proposed procedures can easily carry over to other sensor fusion problems, \eg, lidar-IMU fusion \cite{zuo_lic-fusion_2020}.

\clearpage
\bibliographystyle{splncs04}
\bibliography{bib/zotero}

\begin{thebibliography}{10}
\providecommand{\url}[1]{\texttt{#1}}
\providecommand{\urlprefix}{URL }
\providecommand{\doi}[1]{https://doi.org/#1}

\bibitem{goshen-meskinObservability1992}
{Goshen-Meskin}, D., {Bar-Itzhack}, I.: Observability analysis of piece-wise
  constant systems. {{I}}. {{Theory}}. IEEE Transactions on Aerospace and
  Electronic Systems  \textbf{28}(4),  1056--1067 (Oct 1992).
  \doi{10.1109/7.165367}

\bibitem{hermannNonlinearControllabilityObservability1977}
Hermann, R., Krener, A.: Nonlinear controllability and observability. IEEE
  Transactions on Automatic Control  \textbf{22}(5),  728--740 (Oct 1977).
  \doi{10.1109/TAC.1977.1101601}

\bibitem{heschCameraIMUbasedLocalizationObservability2014}
Hesch, J.A., Kottas, D.G., Bowman, S.L., Roumeliotis, S.I.:
  Camera-{{IMU-based}} localization: {{Observability}} analysis and consistency
  improvement. The International Journal of Robotics Research  \textbf{33}(1),
  182--201 (Jan 2014). \doi{10.1177/0278364913509675}

\bibitem{huaiObservabilityAnalysisKeyframebased2022}
Huai, J., Lin, Y., Zhuang, Y., Toth, C., Chen, D.: Observability analysis and
  keyframe-based filtering for visual inertial odometry with full
  self-calibration. IEEE Transactions on Robotics  (Jan 2022)

\bibitem{isidoriNonlinear2013}
Isidori, A.: Nonlinear {{Control Systems}}. {Springer Science \& Business
  Media} (2013)

\bibitem{jungConstrained2020}
Jung, J.H., Park, C.G.: Constrained filtering-based fusion of images, events,
  and inertial measurements for pose estimation. In: 2020 {{IEEE International
  Conference}} on {{Robotics}} and {{Automation}} ({{ICRA}}). pp. 644--650 (May
  2020). \doi{10.1109/ICRA40945.2020.9197248}

\bibitem{kalmanGeneral1960}
Kalman, R.E.: On the general theory of control systems. In: Proceedings {{First
  International Conference}} on {{Automatic Control}}, {{Moscow}}, {{USSR}}.
  pp. 481--492 (1960)

\bibitem{kellyVisualinertialSensorFusion2011}
Kelly, J., Sukhatme, G.S.: Visual-inertial sensor fusion: {{Localization}},
  mapping and sensor-to-sensor self-calibration. The International Journal of
  Robotics Research  \textbf{30}(1),  56--79 (2011)

\bibitem{li_visual-inertial_2022}
Li, H., Stueckler, J.: Visual-inertial odometry with online calibration of
  velocity-control based kinematic motion models. IEEE Robotics and Automation
  Letters  \textbf{7}(3),  6415--6422 (Jul 2022).
  \doi{10.1109/LRA.2022.3169837}, conference Name: IEEE Robotics and Automation
  Letters

\bibitem{liHighprecisionConsistentEKFbased2013}
Li, M., Mourikis, A.I.: High-precision, consistent {{EKF-based}}
  visual-inertial odometry. The International Journal of Robotics Research
  \textbf{32}(6),  690--711 (May 2013). \doi{10.1177/0278364913481251}

\bibitem{liOnlineTemporalCalibration2014}
Li, M., Mourikis, A.I.: Online temporal calibration for
  camera\textendash{{IMU}} systems: {{Theory}} and algorithms. The
  International Journal of Robotics Research  \textbf{33}(7),  947--964 (2014)

\bibitem{maesObservability2019}
Maes, K., Chatzis, M.N., Lombaert, G.: Observability of nonlinear systems with
  unmeasured inputs. Mechanical Systems and Signal Processing  \textbf{130},
  378--394 (2019)

\bibitem{martinelliVisualinertialStructureMotion2013}
Martinelli, A.: Visual-inertial structure from motion: {{Observability}} and
  resolvability. In: 2013 {{IEEE}}/{{RSJ International Conference}} on
  {{Intelligent Robots}} and {{Systems}} ({{IROS}}). pp. 4235--4242. {IEEE},
  {Tokyo, Japan} (Nov 2013). \doi{10.1109/IROS.2013.6696963}

\bibitem{martinelliObservability2020}
Martinelli, A.: Observability: {{A}} new theory based on the group of
  invariance. Advances in {{Design}} and {{Control}}, {Society for Industrial
  and Applied Mathematics} (Jan 2020). \doi{10.1137/1.9781611976250}

\bibitem{martinelliNonlinearUnknownInput2022}
Martinelli, A.: Nonlinear unknown input observability and unknown input
  reconstruction: {{The}} general analytical solution. arXiv:2201.07610 [cs,
  math]  (Jan 2022)

\bibitem{mirzaeiKalman2008}
Mirzaei, F.M., Roumeliotis, S.I.: A {{Kalman}} filter-based algorithm for
  {{IMU-camera}} calibration: {{Observability}} analysis and performance
  evaluation. IEEE Transactions on Robotics  \textbf{24}(5),  1143--1156 (Oct
  2008). \doi{10.1109/TRO.2008.2004486}

\bibitem{solaQuaternion2017}
Sol{\`a}, J.: Quaternion kinematics for the error-state {{Kalman}} filter.
  Tech. rep., {Institut de Rob\`otica i Inform\`atica Industrial, Barcelona}
  (Nov 2017)

\bibitem{tangINS2009}
Tang, Y., Wu, Y., Wu, M., Wu, W., Hu, X., Shen, L.: {{INS}}/{{GPS}}
  integration: Global observability analysis. IEEE Transactions on Vehicular
  Technology  \textbf{58}(3),  1129--1142 (Mar 2009).
  \doi{10.1109/TVT.2008.926213}

\bibitem{villaverdeInputDependent2019}
Villaverde, A.F., Evans, N.D., Chappell, M.J., Banga, J.R.: Input-dependent
  structural identifiability of nonlinear systems. IEEE Control Systems Letters
   \textbf{3}(2),  272--277 (Apr 2019). \doi{10.1109/LCSYS.2018.2868608}

\bibitem{villaverdeFull2019}
Villaverde, A.F., Tsiantis, N., Banga, J.R.: Full observability and estimation
  of unknown inputs, states and parameters of nonlinear biological models.
  Journal of the Royal Society Interface  \textbf{16}(156),  20190043 (2019)

\bibitem{wuVINS2017}
Wu, K.J., Guo, C.X., Georgiou, G., Roumeliotis, S.I.: {{VINS}} on wheels. In:
  2017 {{IEEE International Conference}} on {{Robotics}} and {{Automation}}
  ({{ICRA}}). pp. 5155--5162 (May 2017). \doi{10.1109/ICRA.2017.7989603}

\bibitem{yangDegenerateMotionAnalysis2019}
Yang, Y., Geneva, P., Eckenhoff, K., Huang, G.: Degenerate motion analysis for
  aided {{INS}} with online spatial and temporal sensor calibration. IEEE
  Robotics and Automation Letters  \textbf{4}(2),  2070--2077 (Apr 2019).
  \doi{10.1109/LRA.2019.2893803}

\bibitem{yangOnlineSelfcalibrationVisualinertial2022}
Yang, Y., Geneva, P., Zuo, X., Guoquan, Huang: Online self-calibration for
  visual-inertial navigation systems: {{Models}}, analysis and degeneracy.
  arXiv:2201.09170 [cs]  (Jan 2022)

\bibitem{zuo_lic-fusion_2020}
Zuo, X., Yang, Y., Geneva, P., Lv, J., Liu, Y., Huang, G., Pollefeys, M.:
  {LIC}-{Fusion} 2.0: {LiDAR}-inertial-camera odometry with sliding-window
  plane-feature tracking. In: 2020 {IEEE}/{RSJ} {International} {Conference} on
  {Intelligent} {Robots} and {Systems} ({IROS}). pp. 5112--5119 (Oct 2020).
  \doi{10.1109/IROS45743.2020.9340704}, iSSN: 2153-0866

\end{thebibliography}

\end{document}